\documentclass[prd,preprintnumbers,nofootinbib]{revtex4} 
\usepackage{graphicx} 
\usepackage{amsmath}
\usepackage{amsfonts,amsbsy}
\usepackage{amssymb}
\usepackage{appendix}

\def\be{\begin{equation}}
\def\ee{\end{equation}}
\def\bea{\begin{eqnarray}}
\def\eea{\end{eqnarray}}

\def\eq#1{{Eq.~(\ref{#1})}}
\def\fig#1{{Fig.~\ref{#1}}}

\begin{document}

\title{\bf  Semi-inclusive photon-hadron production in pp and pA collisions at RHIC and LHC }

%\preprint{}

\author{ Amir H. Rezaeian}
\affiliation{
Departamento de F\'\i sica, Universidad T\'ecnica
Federico Santa Mar\'\i a, Avda. Espa\~na 1680,
Casilla 110-V, Valparaiso, Chile }

\begin{abstract}
We investigate semi-inclusive photon-hadron production in the color glass condensate (CGC)  framework  at RHIC and the LHC energies in  proton-proton (pp) and proton-nucleus (pA)  collisions. We calculate the coincidence probability for azimuthal correlation of pairs of photon-hadron and show that the away-side correlations have a double-peak or a single-peak structure depending on trigger particle selection and kinematics. This novel feature is unique for semi-inclusive photon-hadron production compared to a similar measurement for double inclusive dihadron production in pA collisions. We obtain necessary conditions between kinematics variables for the appearance of a double-peak or a single peak structure for the away-side photon-hadron correlations in pp and pA collisions  at forward rapidities and show that this feature is mainly controlled by the ratio $z_T = p_T^{\text{hadron}}/p_T^{\text{photon}}$. Decorrelation of away-side photon-hadron production by increasing the energy, rapidity and density, and appearance of double-peak structure can be understood by QCD saturation physics. We also provide predictions for the ratio of single inclusive prompt photon to hadron production, and two-dimensional nuclear modification factor for the semi-inclusive photon-hadron pair production at RHIC and the LHC at forward rapidities. 
\end{abstract}

\maketitle

%---------------------------------------------------------------------------
\section{Introduction}

It is generally believed that a system of partons (gluons) at high energy (or small Bjorken-x) forms a new state of matter where the gluon distribution saturates \cite{sg}. Such a system is endowed with a new dynamical momentum scale, the so-called saturation scale at which non-linear gluons recombination effects become as important as the gluon radiation. The color glass condensate (CGC) approach has been proposed to study the physics of gluon saturation at small-x region \cite{mv,cgc-review1}. The CGC formalism is an effective perturbative QCD theory  in which one systematically re-sums quantum corrections which are enhanced by large logarithms of 1/x and also incorporates non-linear high gluon density effects. In the CGC approach, the main features of particle production at high energy remain universal and are controlled by the saturation scale. This picture has been successfully applied to many QCD processes from HERA to RHIC \cite{cgc-review1} and the LHC 
\cite{e-lhc,m1,tr,tr1,j1,me-pa,ridge,ridge0}. In this paper, we will employ the CGC formalism and show that the semi-inclusive photon-hadron ($\gamma-h$) production processes in pA collisions, i.e., $p+A\to \gamma+h+X$, offer more interesting insights to the dynamics of  gluon saturation.   

Photons radiated in hard collisions not via hadronic decays are usually called prompt photon. There are advantages to studying prompt photon production as compared to hadron production. It is theoretically cleaner; one avoids the difficulties involved with description of hadronization and possible initial-state-final-state interference effects which may be present for hadron production and it can be therefore used as a baseline to interpret jet-quenching phenomenon in heavy ion collisions.  A detailed studies of Ref.\,\cite{ja} showed that prompt photon production in pA collisions at RHIC and the LHC at forward rapidities is a sensitive probe of the small-x physics and QCD gluon saturation. On the same line, the semi-inclusive prompt photon-hadron production in pA collisions, has also advantages over a similar production of dihadron. In particular, in dihadron production,  higher number of Wilson lines, and the Weizs\"acker-Williams and the dipole gluon distributions are involved \cite{di} while in the photon-hadron production cross-section, only dipole gluon distribution appears \cite{ja,pho-cgc} which is both experimentally and theoretically well-known, see for example Refs.\,\cite{cgc-review1,bk,bb,nlo,rcbk,jav1,m-b,ipsat}.

Two particle correlations in high-energy collisions have played significant role to reveal QCD novel phenomena \cite{novel,di-e,ph-ex}. 
In particular, the photon-hadron jet correlations have been a very powerful probe of the in-medium parton energy loss in high-energy heavy-ion collisions \cite{ph-ex,ph-th}. It  was suggested in Ref.\,\cite{ja} that the correlation of the back-to-back photon-hadron pair production in high-energy pp and pA collisions can be used to probe the gluon saturation at small-x region and to study the physics of cold nuclear matter in dense region. However, the correlation defined in Ref.\,\cite{ja} may depend crucially on the so-called underlying event and might be rather challenging to measure. Moreover, in Ref.\,\cite{ja} the correlation was studied in a very limited kinematics, see Sec. IV. In this paper, for the first time we study the coincident probability for photon-hadron correlation at RHIC and the LHC in both pp and pA collisions. Dihadron azimuthal angle correlation was already measured by the coincidence probability at RHIC \cite{di-e}. We show that the away-side correlations for a pair of photon-hadron obtained via the coincident probability  have a double or a single peak structure depending on kinematics and whether the trigger particle is selected to be a prompt photon or hadron.  We obtain kinematics conditions for appearance of a double or a single peak structure for the away-side photon-hadron correlations which can be verified by the upcoming experiments at RHIC and the LHC. This novel feature is unique for prompt-hadron production in contrast to dihadron production \cite{di} where the trigger particle can only be a hadron and it was already observed at RHIC \cite{di-e} that the way-side correlation has only single peak structure.  The asymmetric nature of photon-hadron production, and the fact that in semi-inclusive photon-hadron production, QCD and electromagnetic interaction are inextricably intertwined, make the azimuthal correlation of  the produced photon-hadron very intriguing.

We will also provide quantitative predictions for the (two-dimensional) nuclear modification factor for the semi-inclusive photon-hadron production in pA collisions, and the ratio of single inclusive prompt-photon to hadron production in pp and pA collisions, at RHIC and the LHC at forward rapidities. 

This paper is organized as follows; In Sec. II, we first provide a concise description of theoretical framework by introducing the main formulas for the calculation of the cross sections of semi-inclusive photon-hadron (Sec. II-A), single inclusive prompt photon (II-B) and single inclusive hadron (II-C) production within the CGC approach.  In Sec. II-D, we describe how to compute the main ingredient of our formalism, namely dipole-target forward scattering amplitude via the running-coupling Balitsky-Kovchegov evolution equation \cite{bk}. In Sec. III we introduce the observables that we are interested to compute and our numerical setup. In Sec. IV, we present our detailed results and predictions. We summarize our main results in Sec. VI.

%---------------------------------------------------------------------
\section{Theoretical framework}
\subsection{Semi-inclusive prompt photon-hadron production in pp and pA collisions} 

The cross section for production of a prompt photon and a quark with $4$-momenta $p^\gamma$  and $l$ respectively  in scattering of a on-shell quark with $4$-momentum $k$ on a dense target either proton (p) or nucleus (A)  at the leading twist approximation in the CGC formalism is given by  \cite{pho-cgc}, 
\bea
&&{d\sigma^{q\, A \rightarrow q(l)\,\gamma(p^\gamma)\, X}
\over d^2\vec{b_T}\, d^2\vec{p_T}^{\gamma}\, d^2\vec{l_T}\, d\eta_{\gamma}\, d\eta_h} =
{e_q^2\, \alpha_{em} \over \sqrt{2}(4\pi^4)} \, 
{p^-\over  (p_T^\gamma)^ 2 \sqrt{S}} \,
{1 + ({l^-\over k^-})^2 \over
[p^- \, \vec{l_T} - l^- \vec{p_T}^\gamma]^2}\nonumber \\
&&\delta [x_q - {l_T \over \sqrt{S}} e^{\eta_h} - {p_T^\gamma \over \sqrt{S}} e^{\eta_\gamma} ] \,
\bigg[ 2 l^- p^-\, \vec{l_T} \cdot \vec{p_T}^\gamma + p^- (k^- -p^-)\, l_T^2 + l^- (k^- -l^-)\, (p_T^\gamma)^2 \bigg] N_F (|\vec{l_T} + \vec{p_T}^\gamma|,  x_g) ,
\label{cs}
\eea
where $\sqrt{S}$ is the nucleon-nucleon center of mass energy and the light-cone fraction $x_q$ is the ratio of the incoming quark to nucleon energies, namely $x_q =k^-/\sqrt{S/2}$. The pseudo-rapidities of 
outgoing prompt photon $\eta_\gamma$ and quark $\eta_h$ are defined via $p^-={p_T^\gamma \over \sqrt{2}} e^{\eta_{\gamma}}$
and $l^-={l_T \over \sqrt{2}} e^{\eta_h}$.  The subscript $T$ stands for the transverse component. The vector $\vec{b_T}$ denotes the impact-parameter of interaction. The angle between the final-state quark and prompt photon is denoted by $\Delta \phi$ and defined via $\cos(\Delta \phi) \equiv {\vec{l}_T \cdot \vec{p}_T^\gamma \over  l_t p_T^\gamma}$. Throughout this paper, we only consider light hadron production, therefore at high transverse momentum (ignoring hadron mass), the rapidity and pseudo-rapidity is the same. Note that due to the assumption
of collinear fragmentation of a quark into a hadron, the angle $\Delta \phi$ is then the
angle between the produced photon and hadron, assuming that the rapidity of the parent parton and the fragmented hadron is the same. 
In \eq{cs}, $N_F (p_T, x_g)$ is the imaginary part of (quark-antiquark) dipole-target  forward scattering amplitude  which satisfies the Jalilian-Marian-Iancu-McLerran-Weigert-Leonidov-Kovner (JIMWLK) evolution equation \cite{jimwlk,jimwlk1} and has all the multiple scattering and small-$x$ evolution effects encoded (see Sec. II-D). 

In order to relate the above partonic production cross-section to 
proton-target collisions, one needs to convolute the 
partonic cross-section in \eq{cs} with the quark and
antiquark distribution functions of a proton and the quark-hadron 
fragmentation function:
\begin{eqnarray}\label{qh-f}
\frac{d\sigma^{p\, A \rightarrow h (p^h)\, \gamma (p^\gamma)\, X}}{d^2\vec{b_T} \, d^2\vec{p_T}^\gamma\, d^2\vec{p_T}^h \,  
d\eta_{\gamma}\, d\eta_{h}}&=& \int^1_{z_{f}^{min}} \frac{dz_f}{z_f^2} \, 
 \int\, dx_q\,
f _q(x_q,Q^2)  \frac{d\sigma^{q\, A \rightarrow q(l)\,\gamma(p^\gamma)\, X}}
{ d^2\vec{b_T}\, d^2\vec{p_T}^{\gamma}\, d^2\vec{l_T}\, d\eta_{\gamma}\, d\eta_h}  D_{h/q}(z_f,Q^2),
\end{eqnarray}
where $p^h_T$ is the transverse momentum of the produced hadron, and $f_q(x_q,Q^2)$ is the 
parton (quark) distribution function (PDF) of the incoming proton which depends on 
the light-cone momentum fraction $x_q$ and the hard scale $Q$. A summation over the 
quark and antiquark flavors in the above expression should be understood.
The function $D_{h/q}(z_f,Q)$ is the quark-hadron fragmentation function (FF) 
where $z_f$ is the ratio of energies of the produced hadron and quark. 

The light-cone momentum fraction $x_q, x_{\bar q}, x_g$ in Eqs.\,(\ref{cs},\ref{qh-f}) are related to the transverse momenta and
rapidities of the produced hadron and prompt photon via (see appendix in Ref.\,\cite{ja}),
\begin{eqnarray}\label{qh-k}
x_q&=&x_{\bar{q}}=\frac{1}{\sqrt{S}}\left(p_T^\gamma\, e^{\eta_{\gamma}}+\frac{p_T^h}{z_f}\, e^{\eta_h}\right),\nonumber\\
x_g&=&\frac{1}{\sqrt{S}}\left(p_T^\gamma\, e^{-\eta_{\gamma}}+ \frac{p_T^h}{z_f}\, e^{-\eta_{h}}\right),\nonumber\\
z_f&=&p_T^h/l_T, \hspace{1 cm} \text{with}~~~~~ z_{f}^{min}=\frac{p_T^h}{\sqrt{S}}
\left(\frac{e^{\eta_h}}
{1 - {p_T^\gamma\over \sqrt{S}}\, e^{\eta_{\gamma}}}\, 
\right).\label{z_f}\label{ki1}\
\end{eqnarray}

\subsection{Single inclusive prompt photon production in pp and pA collisions} 

The prompt photon cross section in the CGC framework can be readily obtained from
\eq{cs} by integrating over the momenta of the final state quark. After some algebra, the single inclusive prompt photon production can be divided into two contributions of fragmentation and direct photon \cite{ja}: 
\begin{eqnarray}\label{pho2}
\frac{d\sigma^{q\, A \rightarrow \gamma (p^\gamma) \, X}}{d^2 \vec{b_T} d^2 \vec{p_T}^\gamma d\eta_{\gamma}}&=&
\frac{d\sigma^{\text{Fragmentation}}}{d^2 \vec{b_T} d^2 \vec{p_T}^\gamma d\eta_{\gamma}}+\frac{d\sigma^{\text{Direct}}}{d^2 \vec{b_T} d^2 \vec{p_T}^\gamma d\eta_{\gamma}}, 
\\
&=&\frac{1}{(2\pi)^2}\frac{1}{z}\, D_{\gamma/q}(z, Q^2)\, 
N_F(x_g,p_T^\gamma/z) + 
 \frac{e_q^2 \alpha_{em}}{\pi (2\pi)^3}z^2[1+(1 - z)^2]\frac{1}{(p_T^\gamma)^4}
\int_{l_T^2<Q^2}d^2\vec{l_T}\,l_T^2\, 
N_F(\bar{x}_g,l_T), \nonumber\
\end{eqnarray}
where $p^\gamma_T$ is the transverse momentum of the produced  prompt photon, and  $D_{\gamma/q}(z,Q^2)$ is the leading order quark-photon fragmentation function \cite{own}. In order to relate the partonic cross-section given by \eq{pho2} to prompt photon production in pA collisions, we convolute \eq{pho2} with quark and antiquark distribution functions of the projectile proton \cite{me2-pho}, 
 \begin{equation}\label{pho4}
\frac{d\sigma^{p\, A \rightarrow \gamma (p^\gamma) \, X}}{d^2\vec{b_T} d^2\vec{p_T}^\gamma d\eta_{\gamma}}=  
\int_{x_q^{min}}^1 d x_q f_q(x_q, Q^2)
\frac{d\sigma^{q (q^h) \, A \rightarrow \gamma (p^\gamma) \, X}}{d^2 \vec{b_T} d^2\vec{p_T}^\gamma d\eta_{\gamma}},
\end{equation}
where a summation over different quarks (antiquarks) flavors is implicit. The light-cone fraction 
variables $x_g,\bar{x}_g,z$ in Eq.~(\ref{pho2},\ref{pho4})  are related to the transverse momentum of the produced prompt photon and its rapidity \cite{ja}, 
\begin{eqnarray}\label{pho5}
x_g&=& \frac{(p_T^\gamma)^2}{z^2\, x_q\, S} = x_q \, e^{-2\, \eta_\gamma}, \nonumber\\
\bar{x}_g &=& \frac{1}{x_q\, S} \left[{(p_T^\gamma)^2\over z} + \frac{(l_T-p_T^\gamma)^2}{1-z}\right], \nonumber\\
z&\equiv& \frac{p^-}{q^-} = \frac{p_T^\gamma}{x_q\, \sqrt{S}}e^{\eta_{\gamma}} = \frac{x_q^{min}}{x_q}, 
\hspace{1 cm} \text{with}~~~~~ x_q^{min}=z_{min}=\frac{p_T^\gamma}{\sqrt{S}}e^{\eta_{\gamma}}. \
\end{eqnarray}

\subsection{Single inclusive hadron production in pp and pA collisions} 
The cross section for single inclusive hadron production at leading twist approximation, in asymmetric collisions such as pA  or forward rapidity pp collisions at high energy, in the CGC formalism is given by~\cite{dhj,inel},
\begin{equation}\label{final}
\frac{dN^{p A \rightarrow h X}}{d^2\vec{p_T}^h d\eta_h}=\frac{1}{(2\pi)^2}\Bigg[\int_{x_F}^1 \frac{dz}{z^2}\Big[x_1f_g(x_1,Q^2)N_A(x_2,\frac{p_T^h}{z})D_{h/g}(z,Q^2)+\Sigma_qx_1f_q(x_1,Q^2)N_F(x_2,\frac{p_T^h}{z})D_{h/q}(z,Q^2)\Big]+\delta^{\text{inelastic}},
\end{equation}
where the variables $\eta_h$ and $p_T^h$ are the pseudo-rapidity and transverse momentum of the produced hadron. The functions $f_q, N_{F(A)}$ and $D_{h/q}$ in the above are defined the same as in \eq{qh-f}. The indices $q$ and $g$ denote quarks and gluon, with a summation over different flavors being implicit.
The first two terms in the above expersion correspond to elastic contribution, namely an incoming parton scattering elastically  with the CGC target \cite{dhj}. This incoming parton with initial zero transverse momentum picks up transverse momentum of order saturation scale after multiply scattering on the target.  There is also inelastic contribution to the cross-section denoted by $\delta^{\text{inelastic}}$ corresponding to a high transverse momentum parton radiated from the incoming parton in the projectile wave function \cite{inel,me-jamal1}. In this case, the projectile parton interacts with target with small transfer momentum exchanges, but this is enough to decohere the pre-existing high-$p_T$ parton from the hadron wave function and release it as an on-shell particle.  The high-$p_T$ partons in the projectile wave function arise due to DGLAP splitting of partons.
It was shown in Ref.\,\cite{me-jamal1} that at very forward rapidities the inelastic contributions are less important compared to elastic ones  while it is significant at midrapidity at high-energy scatterings.  

The
longitudinal momentum fractions $x_1$ and $x_2$ are defined as follows,
\begin{equation}\label{xs}
x_F\approx \frac{p_T^h}{\sqrt{S}}e^{\eta_h}; \ \ \ \ x_1=\frac{x_F}{ z}; \ \ \ \ \ x_2=x_1e^{-2\eta_h}. 
\end{equation}
One should note that the light-cone fraction variables defined above for the inclusive hadron production are different from the corresponding light-cone variables for single inclusive prompt photon \eq{pho5}  and semi-inclusive photon-hadron  Eq.~(\ref{ki1}) production.

\subsection{Small-x evolution equation and the dipole forward scattering amplitude }
The main ingredient in the cross-section of semi-inclusive photon-hadron production in \eq{cs},  single inclusive prompt photon production in \eq{pho2} and single inclusive hadron production in \eq{final} is the universal dipole forward scattering amplitude which incorporates small-x dynamics and can be calculated via the first-principle non-linear JIMWLK equations \cite{jimwlk,jimwlk1}. In Eqs.\,(\ref{cs},\ref{pho2},\ref{final}), the amplitude $N_F$ ($N_A$) is the two-dimensional Fourier
transformed of the imaginary part of the forward dipole-target
scattering amplitude $\mathcal{N}_{A(F)}$ in the fundamental (F) or adjoint (A)
representation,
\begin{equation}
N_{A(F)}(x,k_T)=\int d^2\vec r e^{-i\vec k_T.\vec r}\left(1-\mathcal{N}_{A(F)}(r,Y=\ln(x_0/x))\right),
\end{equation}
where $r=|\vec r|$ is the dipole transverse size. In the large-$N_c$
limit, one has the following relation between the adjoint and fundamental dipoles, 
\begin{equation}
\mathcal{N}_A(r,Y)=2\mathcal{N}_F(r,Y)-\mathcal{N}_F^2(r,Y). 
\end{equation}
In the large $N_c$ limit, the coupled JIMWLK equations are simplified to the Balitsky-Kovchegov (BK) equation \cite{bk,bb,nlo,rcbk}, a closed-form equation for the rapidity evolution of the dipole amplitude in which both linear radiative processes and non-linear recombination effects are systematically incorporated. The running-coupling BK (rcBK) equation  has the following simple form:
\begin{equation}
  \frac{\partial\mathcal{N}_{A(F)}(r,x)}{\partial\ln(x_0/x)}=\int d^2{\vec r_1}\
  K^{{\rm run}}({\vec r},{\vec r_1},{\vec r_2})
  \left[\mathcal{N}_{A(F)}(r_1,x)+\mathcal{N}_{A(F)}(r_2,x)
-\mathcal{N}_{A(F)}(r,x)-\mathcal{N}_{A(F)}(r_1,x)\,\mathcal{N}_{A(F)}(r_2,x)\right]\,,
\label{bk1}
\end{equation}
where the evolution kernel $K^{{\rm run}}$ using Balitsky`s
prescription \cite{bb} for the running coupling is defined as,
\begin{equation}
  K^{{\rm run}}(\vec r,\vec r_1,\vec r_2)=\frac{N_c\,\alpha_s(r^2)}{2\pi^2}
  \left[\frac{1}{r_1^2}\left(\frac{\alpha_s(r_1^2)}{\alpha_s(r_2^2)}-1\right)+
    \frac{r^2}{r_1^2\,r_2^2}+\frac{1}{r_2^2}\left(\frac{\alpha_s(r_2^2)}{\alpha_s(r_1^2)}-1\right) \right],
\label{kbal}
\end{equation}
with $\vec r_2 \equiv \vec r-\vec r_1$. The only external input for the rcBK non-linear equation is the initial condition for the evolution which is taken to have the following form motivated by McLerran-Venugopalan (MV) model \cite{mv},  
  \begin{equation}
\mathcal{N}(r,Y\!=\!0)=
1-\exp\left[-\frac{\left(r^2\,Q_{0s}^2\right)^{\gamma}}{4}\,
  \ln\left(\frac{1}{\Lambda\,r}+e\right)\right].
\label{mv}
\end{equation}
where the infrared scale is taken $\Lambda=0.241$ GeV and the onset of small-x evolution is assumed to be at
$x_0=0.01$ \cite{jav1}. The only free parameters in the above are $\gamma$ and the initial saturation scale $Q_{0s}$ (probed by quarks), with a notation $s=p$ and, $A$ for a proton and nuclear target, respectively. The initial saturation scale of proton $Q_{0p}^2\simeq 0.168\,\text{GeV}^2$  with the corresponding $\gamma \simeq 1.119$ were extracted from a global fit to proton structure functions in DIS in the small-x region \cite{jav1} and single inclusive hadron data in pp collisions at RHIC and the LHC  \cite{j1,me-jamal1,jm,raj}.  Note that the current HERA data alone is not enough to uniquely fix the values of $Q_{0p}$ and $\gamma$ \cite{jav1}. The recent LHC data, however, seems to indicate that $\gamma>1$ is preferable \cite{j1}.  We will consider the uncertainties coming from our freedom to choose among different parameter sets for the rcBK description of the proton.

Notice that in the rcBK equation \eq{bk1}, the impact-parameter dependence of the collisions was
ignored. Solving the rcBK equation in the presence of the impact-parameter is still open problem \cite{bk-b}. However, for the minimum-bias analysis considered here this may not be important. Then, the initial saturation scale of a nucleus $Q_{0A}$ should be considered as an impact-parameter averaged value and it is extracted from the minimum-bias data.  For the minimum-bias collisions, one may assume that the initial saturation scale of a nucleus with atomic mass number A, scales linearly with $A^{1/3}$ \cite{mv}, namely we have $Q_{0A}^2=cA^{1/3}~Q_{0p}^2$ where the parameter $c$ is fixed from a fit to data\footnote{Note that a different $A$-dependence of the nuclear saturation scale with a pre-factor fitted to the HERA data, numerically leads to a very similar relation between the proton and nuclear saturation scale \cite{urs}.}. In Ref.\,\cite{raj}, it was shown that  DIS data for heavy nuclear targets can be described with $c\approx 0.5$. This is consistent with the fact that RHIC inclusive hadron production data in minimum-bias deuteron-gold collisions prefers an initial saturation scale within $Q_{0A}^2\approx 3 \div 4~Q_{0p}^2$  \cite{me-jamal1,jm}. We will take into account the uncertainties associated to the variation of initial saturation scale in the rcBK evolution equation. 

\section{observables and numerical setup}

In this paper, we only consider observables which are defined as a ratio of cross sections.  We expect that some of the theoretical uncertainties, such as sensitivity to $K$ factors which effectively incorporates the missing higher order corrections, will drop out in this way. Therefore, we take $K=1$ throughout this paper.
We start by considering the ratio of inclusive prompt photon to the neutral pion production, defined as
\begin{equation}\label{ratio-1}
\frac{\gamma^{inclusive}}{\pi^0} (p_T^\gamma,\,p_T^h;\,\eta_h,\,\eta_\gamma)=\frac{dN^{p\, A \rightarrow \gamma(p_T^\gamma)  \, X}}{d^2\vec{p_T}^\gamma d\eta_{\gamma}} /
 \frac{dN^{p A \rightarrow h(p_T^h) X}}{d^2\vec{p_T}^h d\eta_h},
\end{equation} 
where the cross-section for the single inclusive prompt photon and hadron production in pA and pp  collisions are given in Eqs.\,(\ref{pho4},\ref{final}).

In order to investigate the azimuthal angle correlations between the produced prompt photon and hadron, we calculate the coincidence probability. In the contrast to a more symmetric production like dihadron, for the photon-hadron production we have freedom to select the trigger particle to be a produced prompt photon or a hadron. We consider here both cases. In a case that the trigger particle is a prompt photon,  the coincidence probability is defined as $CP_h(\Delta \phi)=N^{\text{pair}}_h (\Delta \phi)/N_{\text{photon}}$, where $N^{\text{pair}}_h (\Delta \phi)$ is the yield of photon-hadron pair production including a associated hadron (neutral pion) with a transverse momentum $p^h_{T,S}$ with a trigger (leading) prompt-photon with transverse momentum $p^\gamma_{T,L}$  and the azimuthal angle between them  $\Delta \phi$. In the same fashion, one can define the coincidence probability with hadron-triggered as $CP_\gamma(\Delta \phi)=N^{\text{pair}}_\gamma (\Delta \phi)/N_{\text{hadron}}$ where $N^{\text{pair}}_\gamma (\Delta \phi)$ is the yield of photon-hadron pair including a associated prompt photon and a trigger hadron (neutral pion) with transverse momentum denoted by  $p^\gamma_{T,S}$ and $p^h_{T,L}$, respectively,   
\begin{eqnarray} 
CP_{h}(\Delta \phi;  p^h_{T,S}, p^\gamma_{T,L}; \eta_\gamma,\eta_h)&=&\frac{2\pi \int_{p^\gamma_{T,L}} dp_T^\gamma p_T^\gamma \int_{p^h_{T,S}} dp_T^h p_T^h\frac{dN^{p\, A \rightarrow h(p_T^h)\,\gamma(p_T^\gamma)\, X} }
{d^2\vec{p_T}^\gamma\, d^2\vec{p_T}^h\, d\eta^{\gamma}\, d\eta^h }}
 {\int_{p^\gamma_{T,L}} d^2\vec{p_T}^\gamma\, \frac{dN^{p\, A \rightarrow \gamma(p_T^\gamma)\, X}}
{ d^2\vec{p_T}^\gamma\,d\eta_{\gamma} }}, \label{cp1}\\
CP_{\gamma}(\Delta \phi; p^\gamma_{T,S}, p^h_{T,L}; \eta_\gamma,\eta_h)&=&\frac{2\pi \int_{p^h_{T,L}} dp_T^h p_T^h  \int_{p^\gamma_{T,S}} dp_T^\gamma p_T^\gamma\frac{dN^{p\, A \rightarrow h(p_T^h)\,\gamma(p_T^\gamma)\, X} }
{d^2\vec{p_T}^\gamma\, d^2\vec{p_T}^h\, d\eta^{\gamma}\, d\eta^h }}
 {\int_{p^h_{T,L}} d^2\vec{p_T}^h\, \frac{dN^{p\, A \rightarrow h(p_T^h)\, X}}
{ d^2\vec{p_T}^h\,d\eta_h }},\label{cp2}\
\end{eqnarray}
where the integrals are performed within given momenta bins denoted by subscript $p^\gamma_{T,L}, p^\gamma_{T,S}, p^h_{T,L}$ and  $p^h_{T,S}$.  The yields in the above expersion are defined in Eqs.\,(\ref{cs},\ref{pho2},\ref{final}). Similar to the dihadron correlation measurements at RHIC \cite{di-e}, in the definition of the coincidence probability, we follow a convention that a leading or trigger particle has transverse momentum {\it larger} than an associated particle. Later, we will also study, the implication of different kinematics definitions for the trigger particle in $\gamma-h$ correlations.  
\begin{figure}[t]                                                            
                                  \includegraphics[width=7.5 cm] {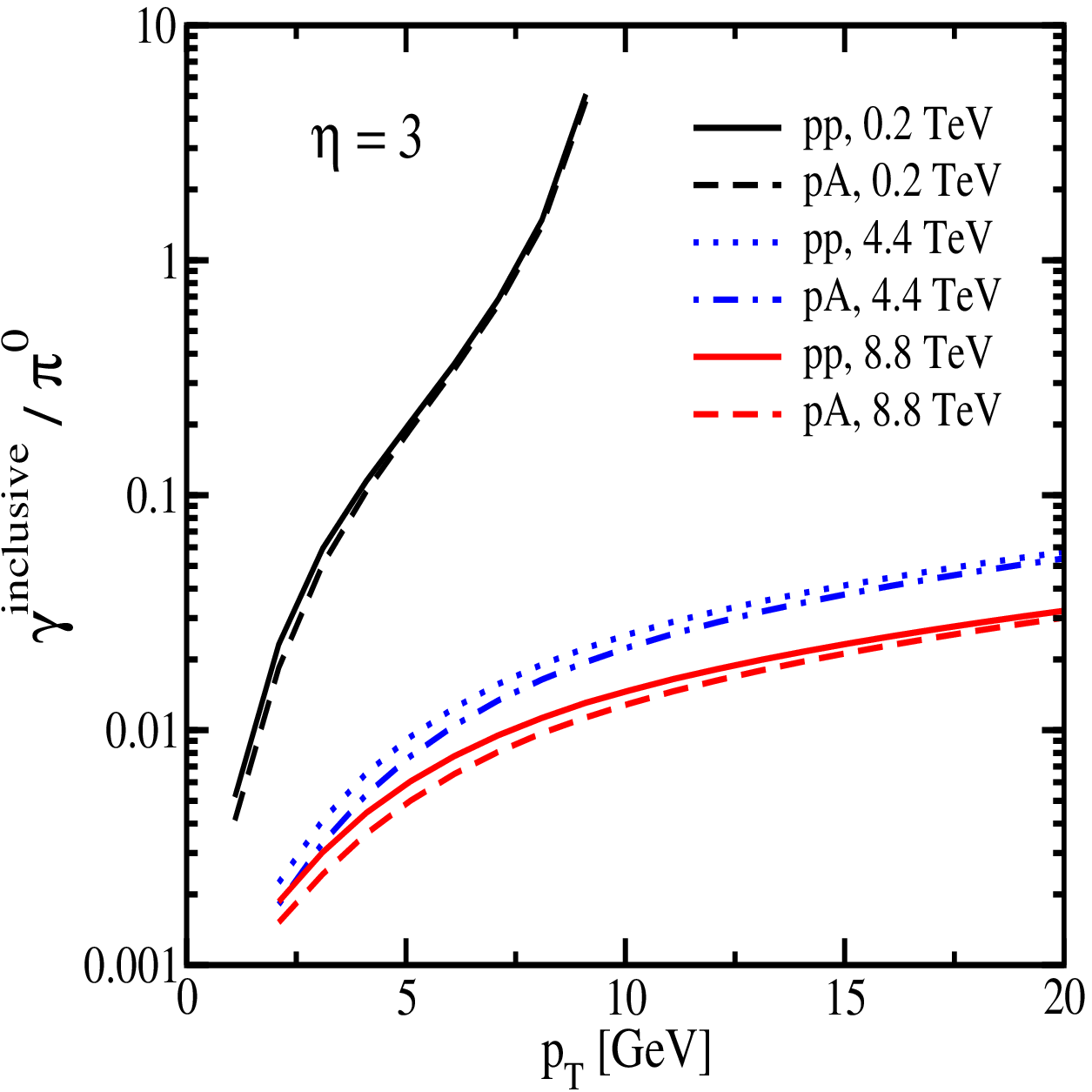}   
                                  \includegraphics[width=7.5 cm] {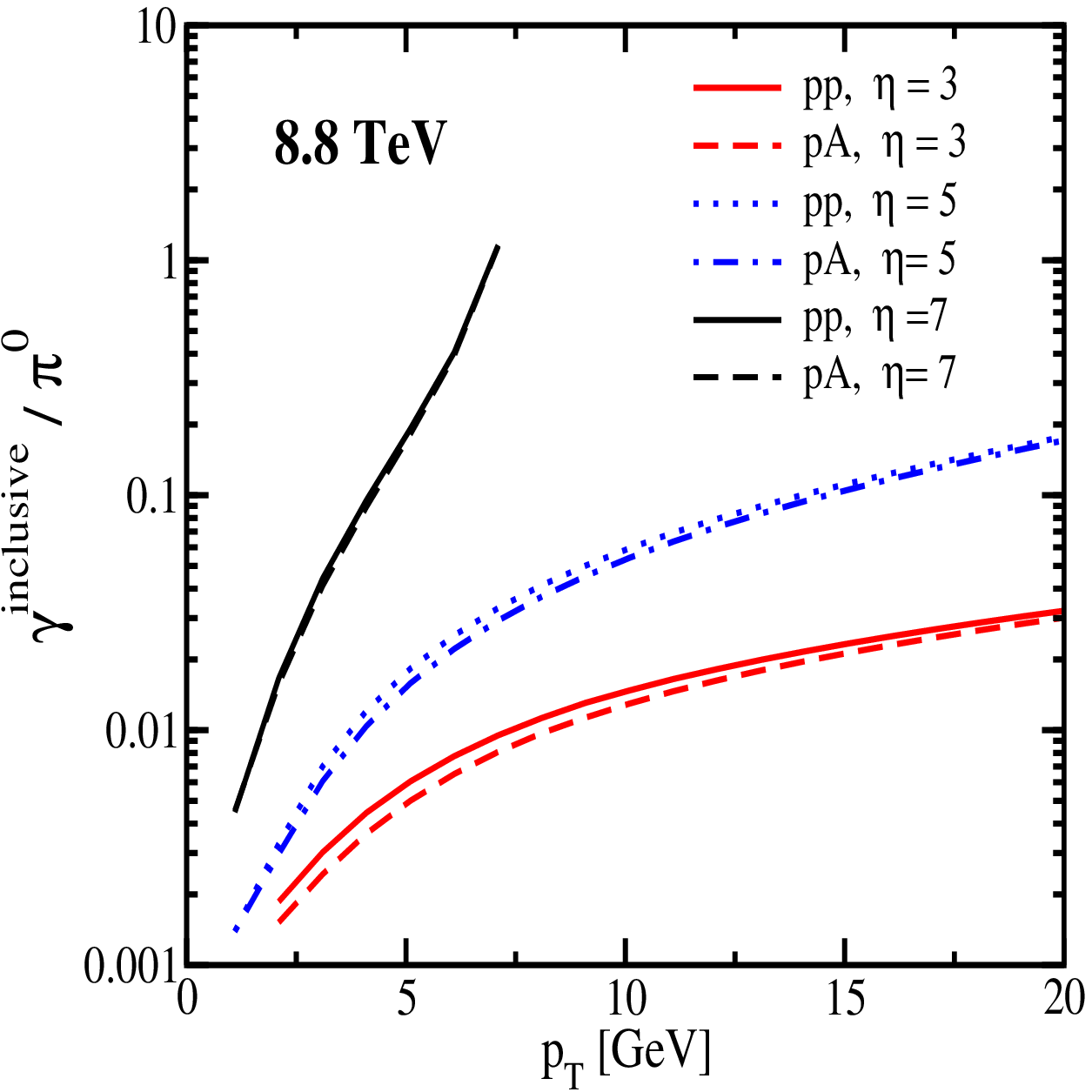}       
\caption{The ratio $\gamma^{inclusive}/\pi^0$ as a function of transverse momentum $p^\gamma_T=p^h_T=p_T$ at various rapidities $\eta_h=\eta_{\gamma}=\eta$ and energies in minimum bias pp and pA collisions.   }
\label{fig-r}
\end{figure}

In nuclear collisions, nuclear effects on particle production  may be evaluated in terms of ratios of particle yields in pA and pp collisions (scaled with a proper normalization), the so-called  nuclear modification factor $R_{pA}$. The nuclear modification factor for semi-inclusive photon-hadron production is defined as,
\begin{eqnarray}
R_{pA}^{h\gamma } ( \Delta \phi;  p^h_T, p^\gamma_T; \eta_\gamma,\eta_h)&=&\frac{1}{N_{coll}} 
\frac{dN^{p\, A \rightarrow h(p^h_T)\,\gamma(p^{\gamma}_T)\, X} } {d^2\vec{p_T}^h\, d^2\vec{p_T}^\gamma\, d\eta_{\gamma}\, d\eta_h}
/\frac{dN^{p\, p \rightarrow h(p^h_T)\,\gamma(p^{\gamma}_T)\, X} } {d^2\vec{p_T}^h\, d^2\vec{p_T}^\gamma\, d\eta_{\gamma}\, d\eta_h},\label{rp1}\\
R_{pA}^{h\gamma } ( p^h_T, p^\gamma_T; \eta_\gamma,\eta_h)&=&\frac{1}{N_{coll}} 
\frac{dN^{p\, A \rightarrow h(p^h_T)\,\gamma(p^{\gamma}_T)\, X} } {dp_T^h\, dp_T^\gamma\, d\eta_{\gamma}\, d\eta_h}
/\frac{dN^{p\, p \rightarrow h(p^h_T)\,\gamma(p^{\gamma}_T)\, X} } {dp_T^h\, dp_T^\gamma\, d\eta_{\gamma}\, d\eta_h}, \label{rp2}
\end{eqnarray}
where the photon-hadron yield  in high-energy pA and pp collisions is given in \eq{cs}. In \eq{rp2}, the integrals over the angles were performed. The normalization constant $N_{coll}$ is the number of binary proton-nucleus collisions. We take $N_{coll}=3.6$ and  $7.4$ at $\sqrt{s}=0.2$ and $8.8$ TeV, respectively, in pA collisions  \cite{ncoll}. Notice that in our approach $N_{coll}$ is taken from outset and one should take into account possible discrepancy between our assumed normalization $N_{coll}$ and the experimentally measured value for $N_{coll}$ by rescaling our curves.

We will use the NLO MSTW 2008 PDFs \cite{mstw} and the 
NLO KKP FFs \cite{kkp} for neutral pion. For the photon fragmentation function, we will use the full leading log parametrization \cite{own,ffp}.  We assume the factorization scale $Q$ in the FFs and the PDFs to be equal and its value is taken to be  $p_T^h$ and $p_T^\gamma$ for inclusive (and semi-inclusive) hadron and prompt photon production, respectively.

\begin{figure}[t]       
                              \includegraphics[width=7.5 cm] {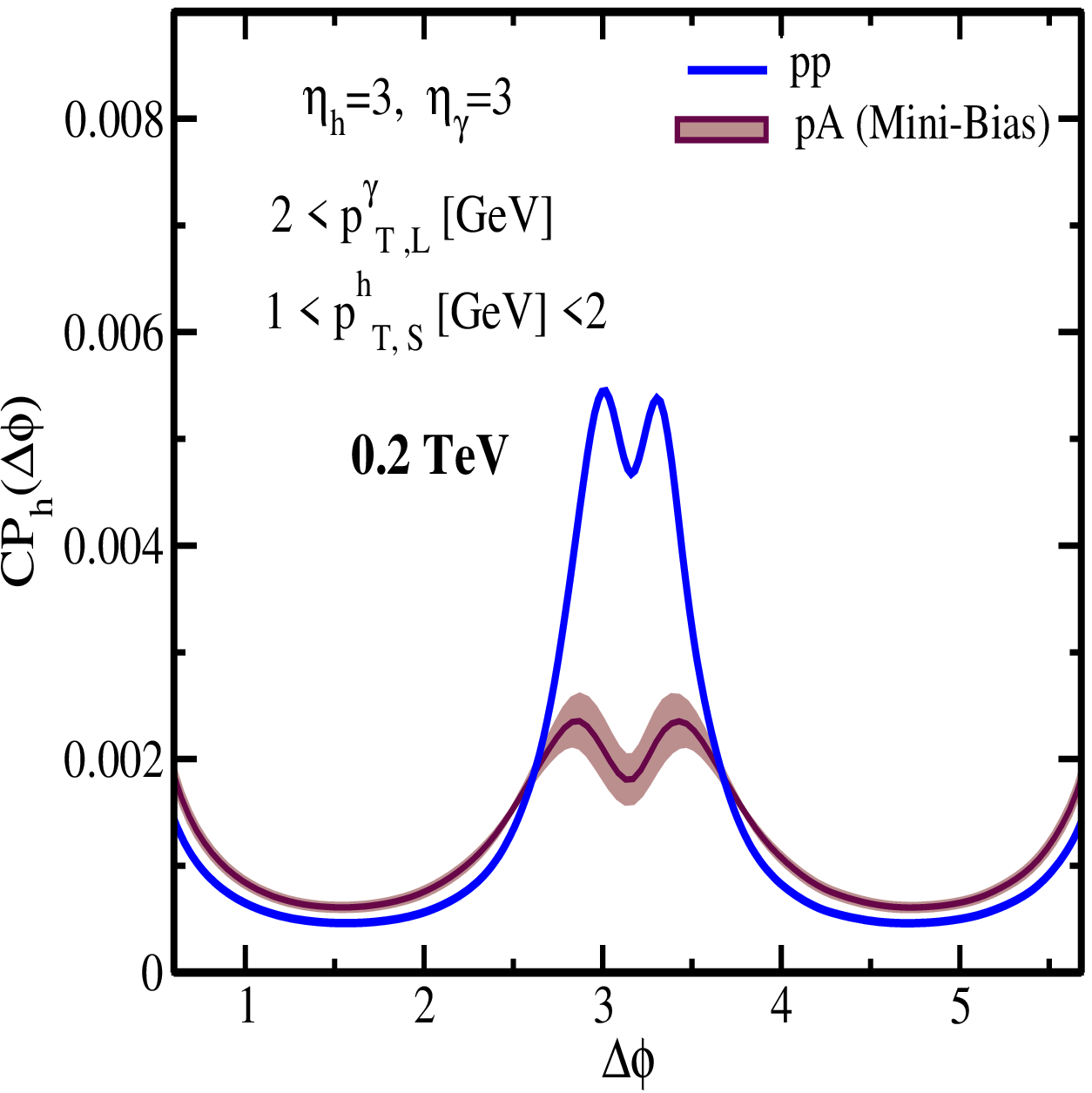}  
                                  \includegraphics[width=7.5 cm] {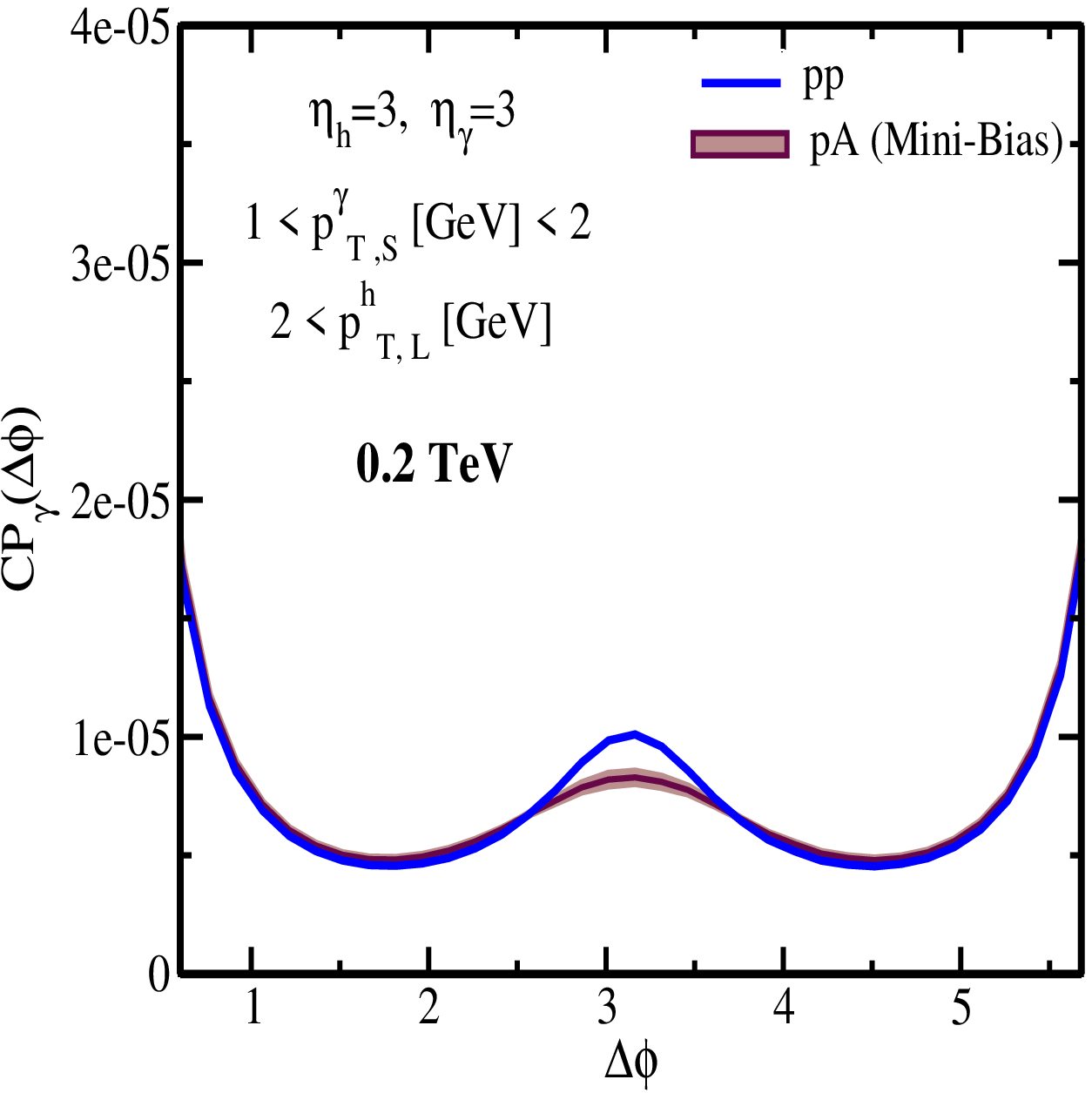}                                 
                                 \includegraphics[width=7.5 cm] {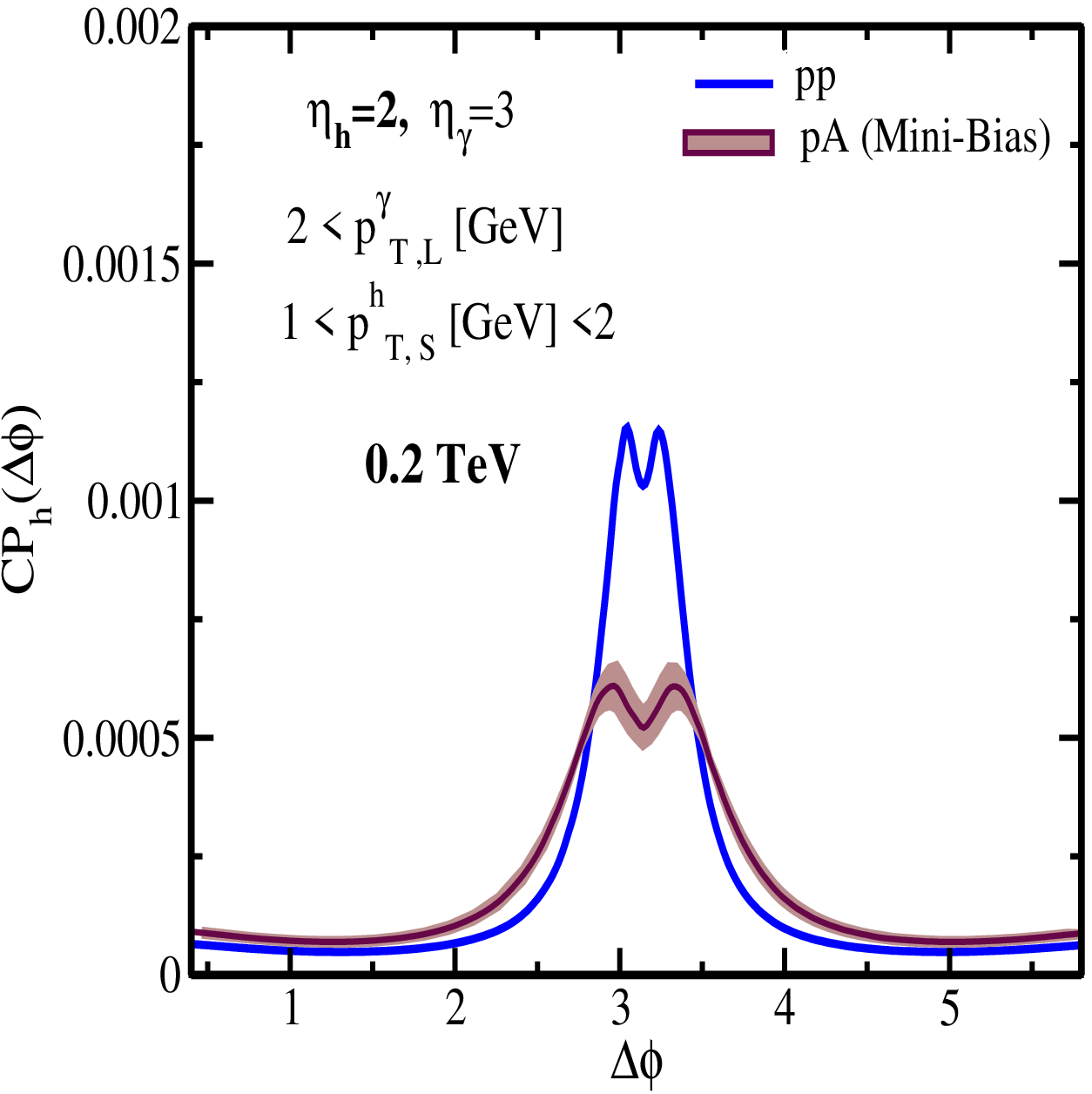}   
\caption{The photon-hadron ($\gamma-\pi^0$) azimuthal correlation (the coincidence probability) $CP_{h}(\Delta \phi)$ and $CP_{\gamma}(\Delta \phi)$ defined in  Eqs.\,(\ref{cp1},\ref{cp2}) in minimum-bias (Mini-Bias) pA and pp collisions at RHIC $\sqrt{S}=0.2$ TeV at different rapidities for the produced hadron $\eta_h$ and inclusive prompt photon $\eta_\gamma$. In the plot, the values of transverse momenta bins of the associated (and leading) neutral pion $p^h_{T,S}$ (and $p^h_{T,L}$) and leading (and associated) prompt photon  $p^{\gamma}_{T,L}$  (and $p^{\gamma}_{T,S}$) are given.   }
\label{fig1}
\end{figure}
%%%%%%%%%%%%%%%%%%%%%%%%%%%%%%%%
%%%%%%%%%%%%%%%%%%%%%%%%%%%%%%%%
\section{Main results and predictions}

In \fig{fig-r} (right), we show the ratio of single inclusive prompt photon to neutral pion ($\pi^0$) production defined via \eq{ratio-1}  for $\eta_h=\eta_{\gamma}=\eta$ and $p^\gamma_T=p^h_T=p_T$  at the LHC energy $\sqrt{S}=8.8$ TeV  as a function of  transverse  momentum $p_T$ in minimum bias pp and pA collisions at different rapidities $\eta$. It is seen that at the LHC, the ratio $\gamma^{inclusive}/\pi^0$ is smaller than one for a large range of rapidities.  In \fig{fig-r} (left), we compare the ratio $\gamma^{inclusive}/\pi^0$ at a fixed rapidity 
$\eta=3$ but different energies. In our approach, a fast valence quark from the projectile proton radiates a photon before and after multiply interaction on the color-glass-condensate target \cite{ja}. The prompt photon can be mainly produced  by quark (at the leading log approximation), while pions can be produced by both projectile gluons and quarks, see Eqs.\,(\ref{pho4},\ref{final}). At the LHC energies at around midrapidity, gluons dominate and consequently the pion production rate is higher than prompt photon while for forward collisions and hight $p_T$ we have $x_1\to 1$, therefore {\it projectile} quarks distributions enhance and consequently the prompt photon production rate grows with increasing rapidity. This can be seen from \fig{fig-r}, namely the ratio $\gamma^{inclusive}/\pi^0 $ increases with rapidity and transverse momentum while it decreases with energy.  Note that in our picture, the description of the target appears via the dipole-target forward scattering amplitude  and it numerically drops out in the ratio, and as a consequence  the ratio $\gamma^{inclusive}/\pi^0 $ is approximately equal for pp and pA collisions at high $p_T$ and is not sensitive to the initial saturation scale\footnote{Note that  in calculation of the ratio of $\gamma^{inclusive}/\pi^0$ we ignored the inelastic contributions in both inclusive prompt photon and hadron production cross-sections assuming that higher order terms will be canceled out in the ratio. }.

\begin{figure}[t]       
                               \includegraphics[width=7.5 cm] {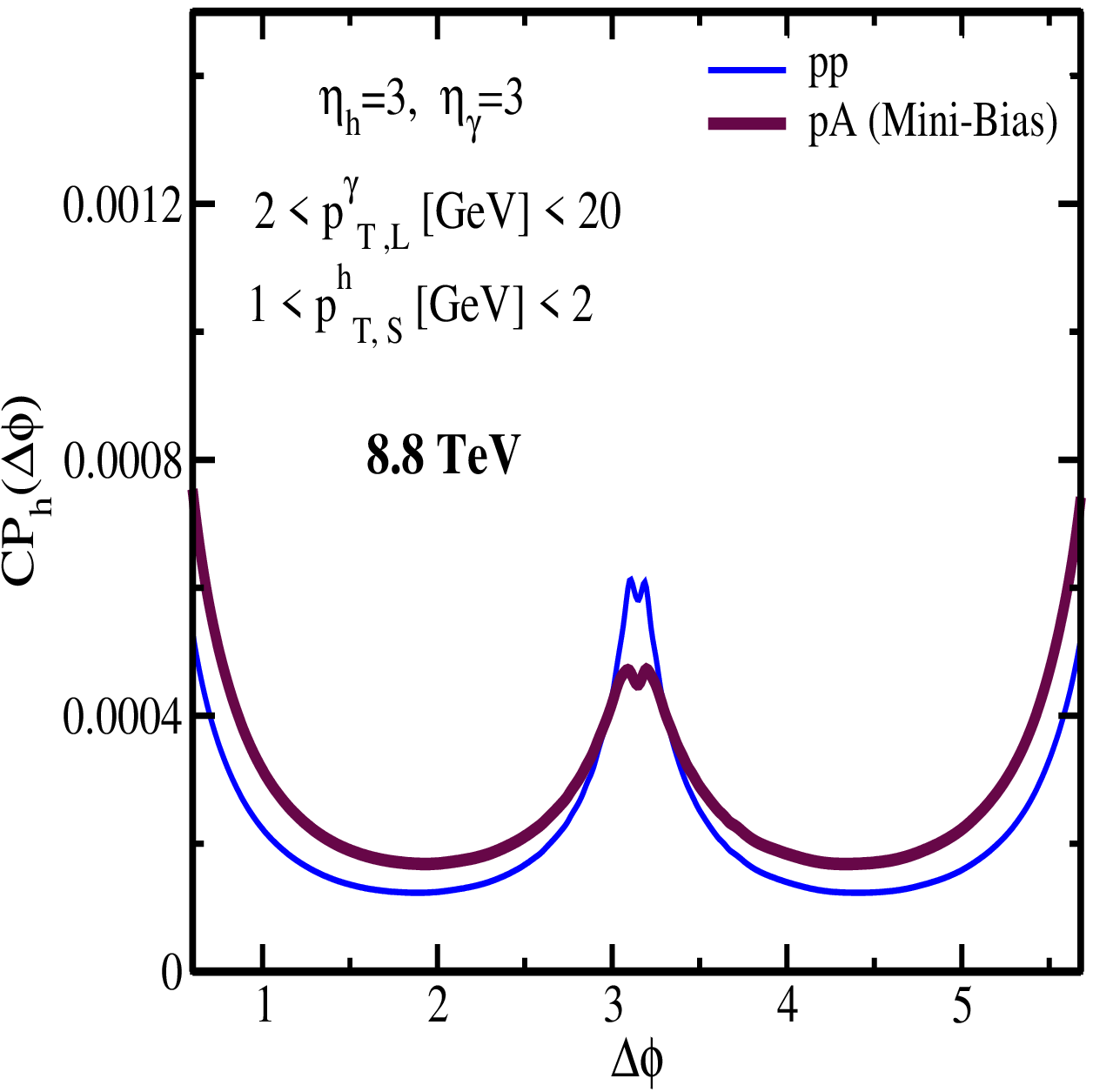}  
                               \includegraphics[width=7.5 cm] {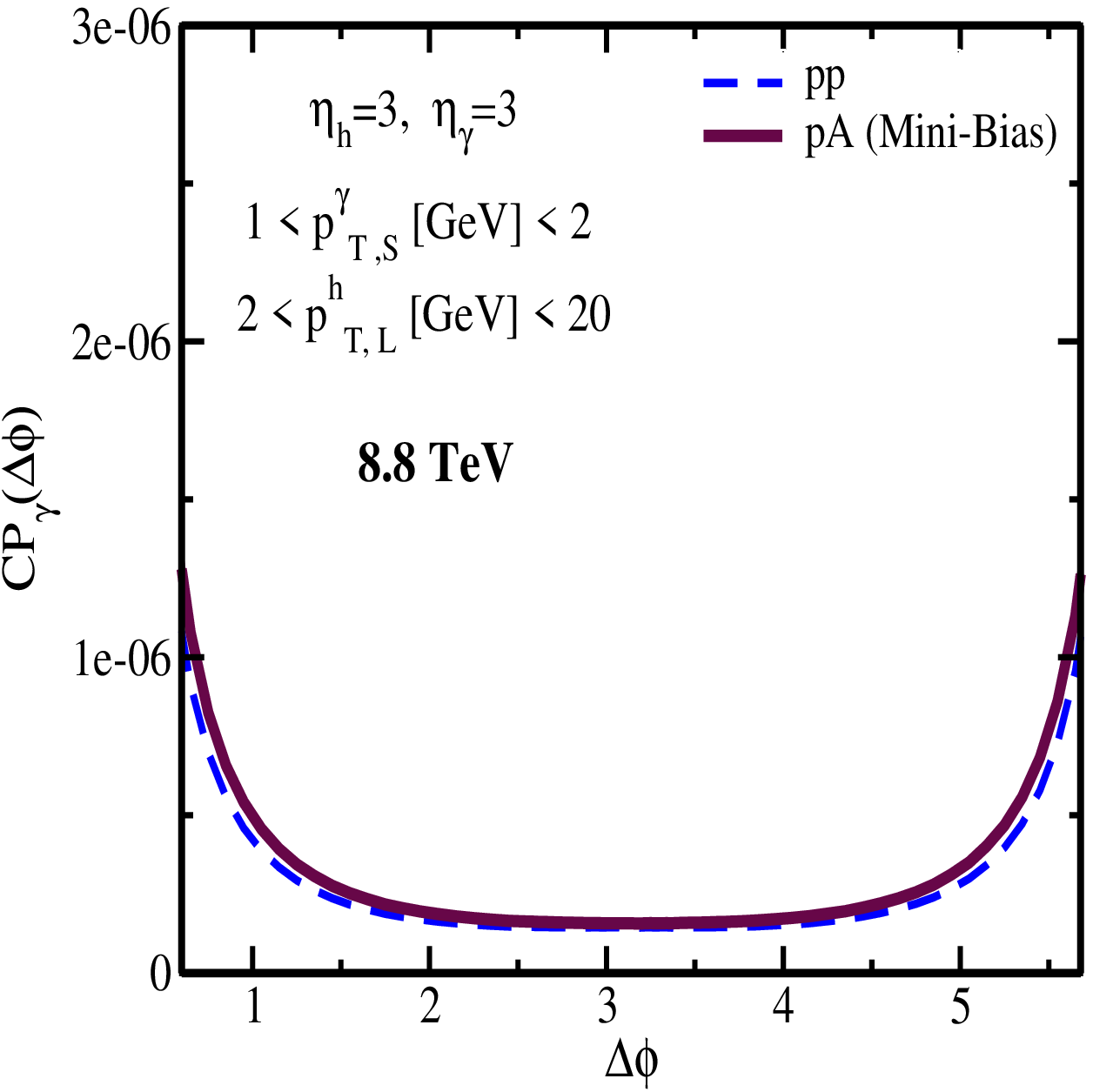} 
                              \includegraphics[width=7.5 cm] {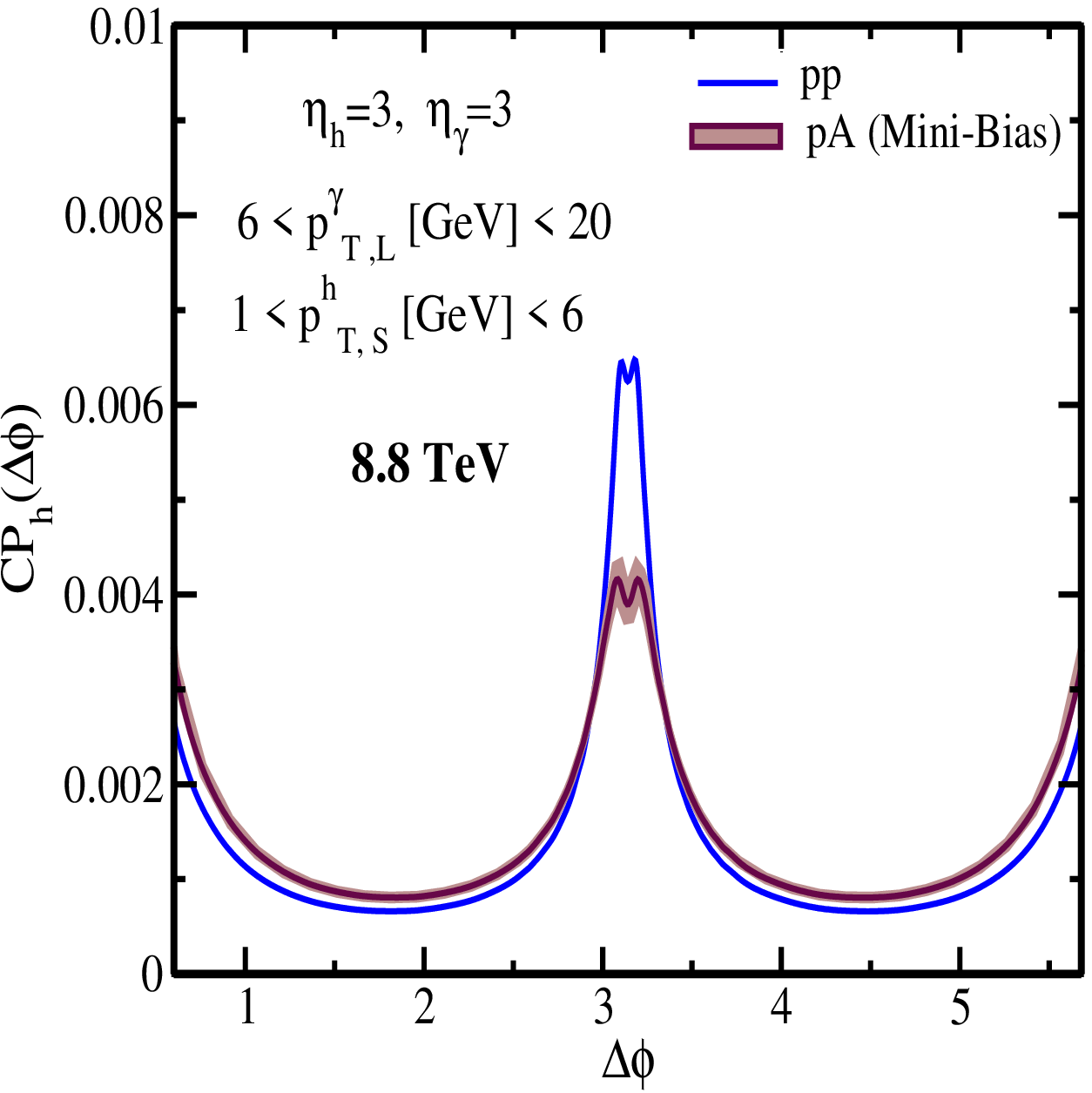}                                     
                                 \includegraphics[width=7.5 cm] {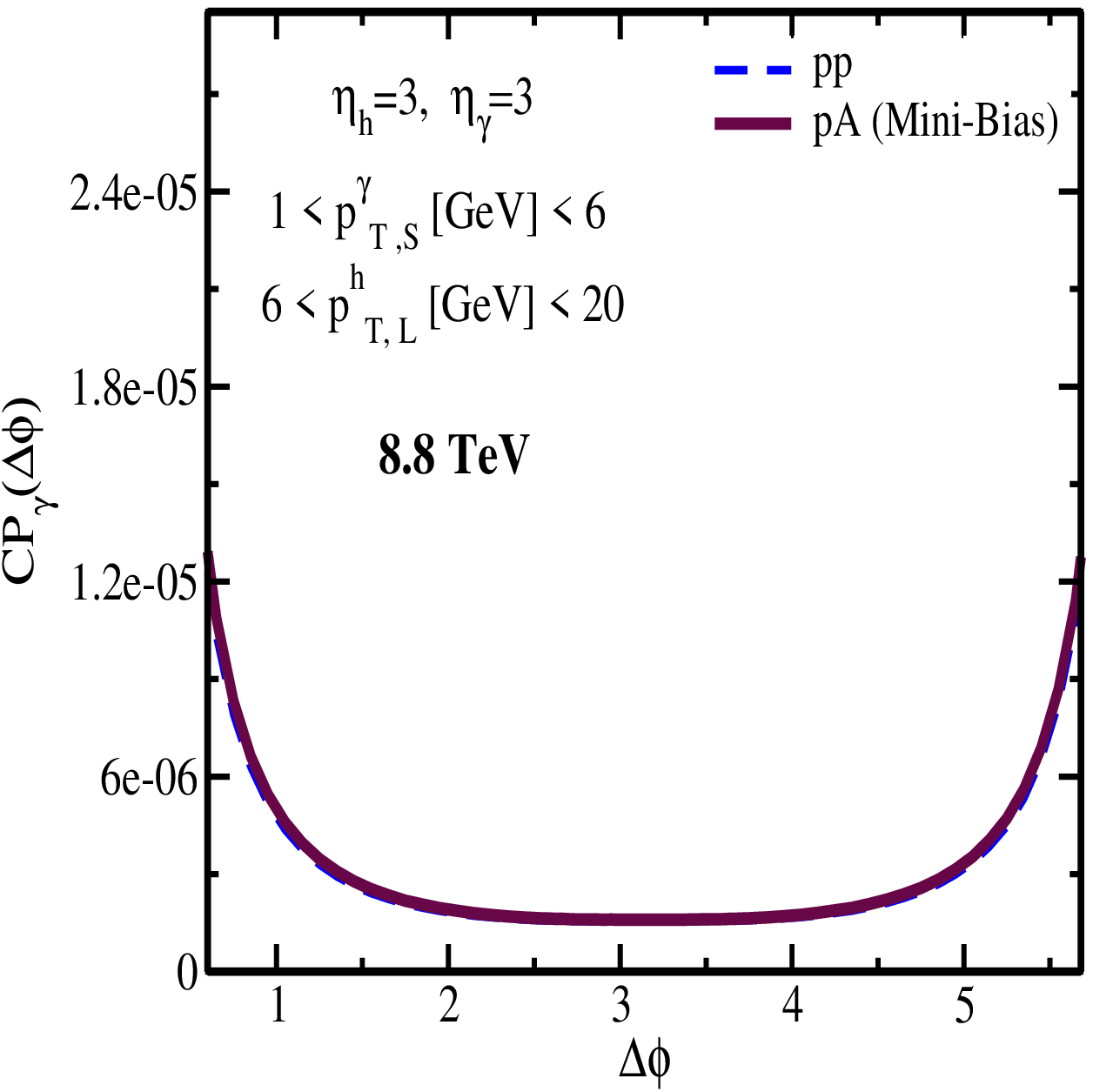}                                      
\caption{ The $\gamma-\pi^0$ coincidence probability $CP_h(\Delta \phi)$ and $CP_{\gamma}(\Delta \phi)$  in minimum-bias pA  and pp collisions at the LHC $\sqrt{S}=8.8$ TeV at $\eta_h=\eta_{\gamma}=3$ for two bins of transverse momenta of the associated (and leading) neutral pion $p^h_{T,S}$ (and $p^h_{T,L}$) and leading (and associated) prompt photon  $p^{\gamma}_{T,L}$  (and $p^{\gamma}_{T,S}$). }
\label{fig2}
\end{figure}

\begin{figure}[t]       
                               \includegraphics[width=7.5 cm] {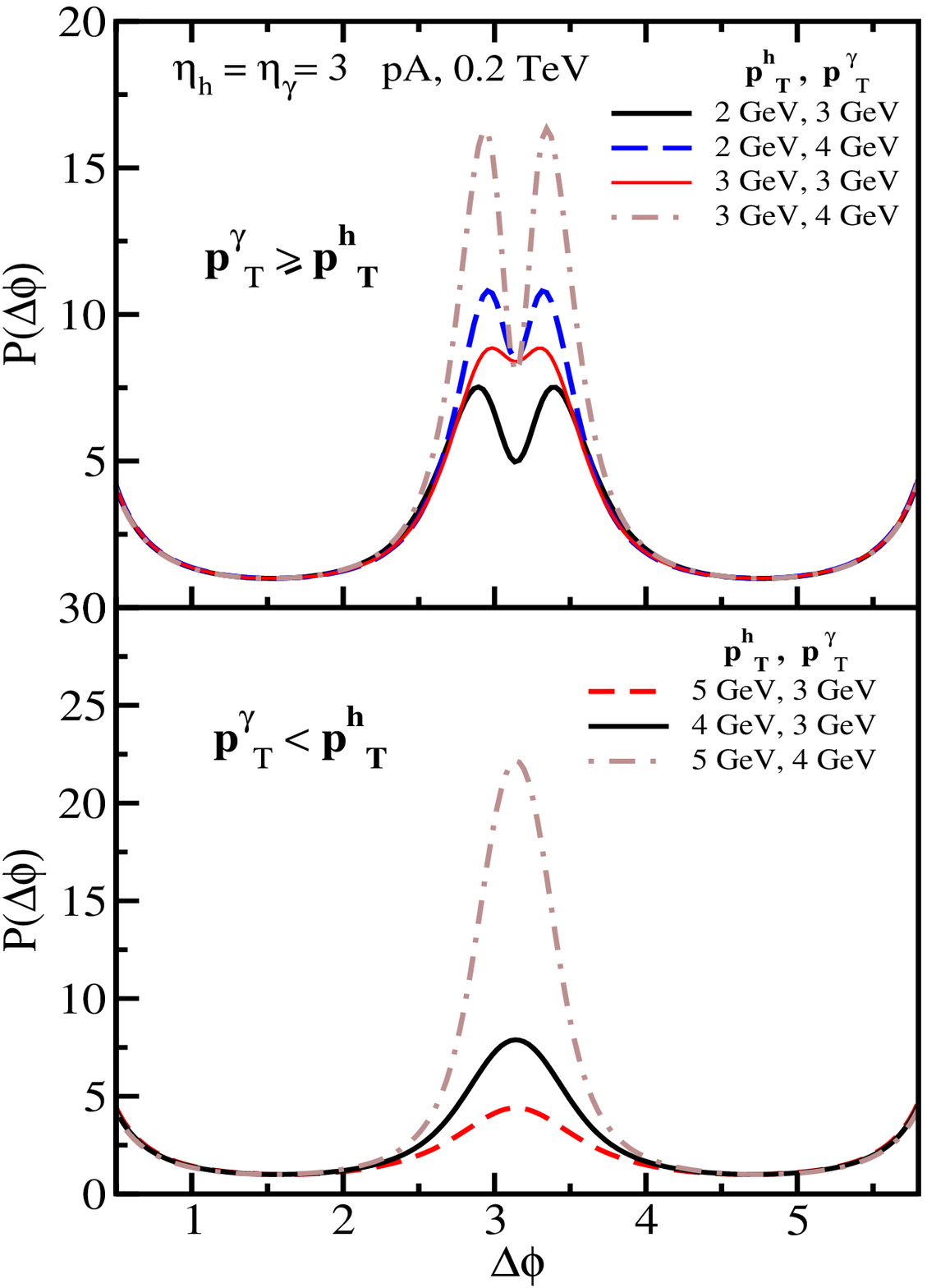}  
                               \includegraphics[width=7.5 cm] {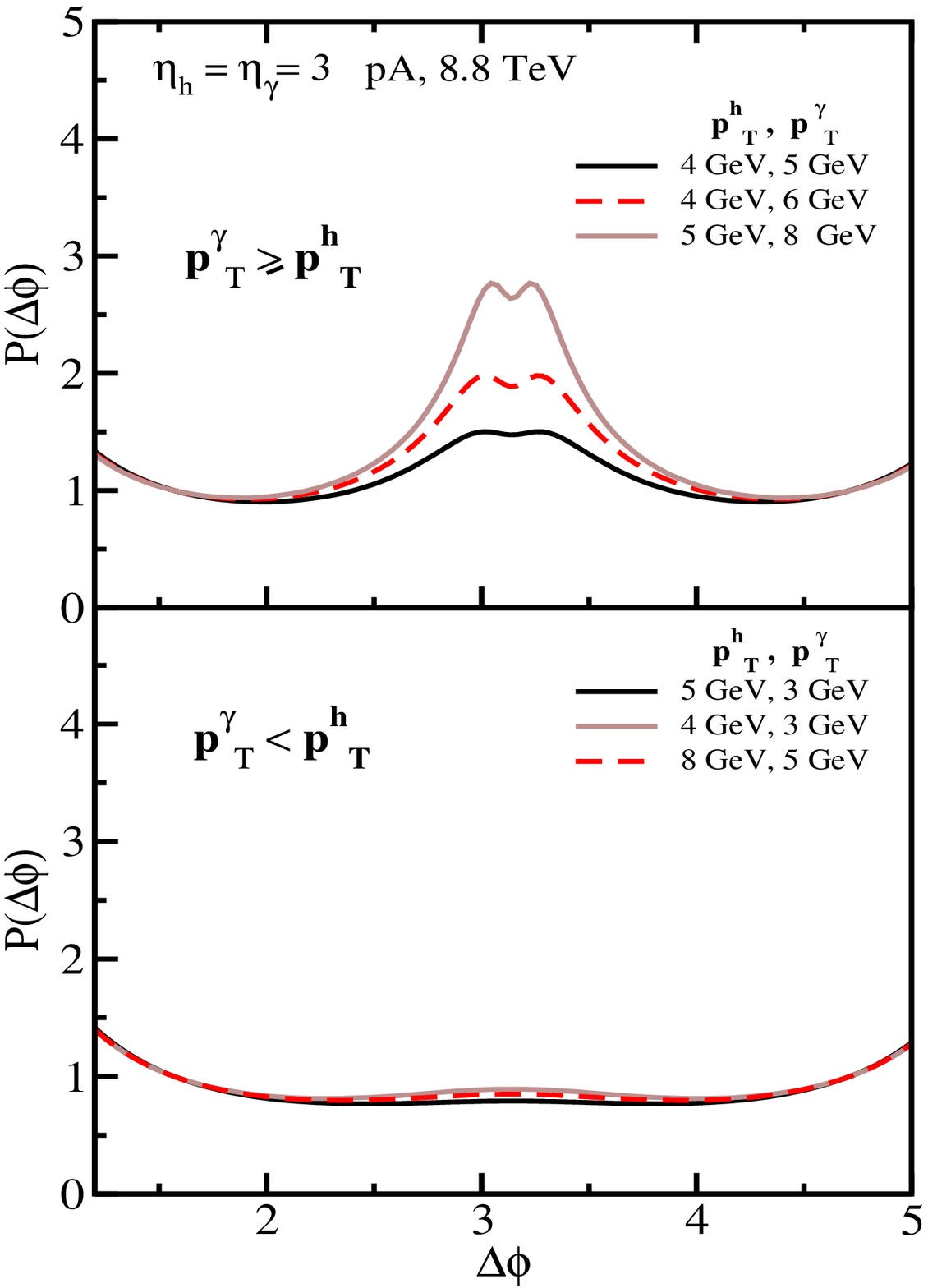}                              
\caption{ The photon-hadron correlation $P(\Delta \phi)$ defined in \eq{az}  in minimum-bias pA  collisions at the LHC (right) and RHIC (left)  at forward rapidity for various transverse momenta of produced prompt photon $p_T^\gamma$ and hadron $p_T^h$ corresponding to two cases of $z_T>1$ and $z_T<1$.  }
\label{fig2-pt}
\end{figure}
Next, we study the azimuthal angle correlation of photon-hadron ($\gamma-\pi^0$) production by computing the coincidence probability defined in Eq.\,(\ref{cp1},\ref{cp2}).  
In Figs.\,(\ref{fig1},\ref{fig2})  we show the coincidence probability for photon-hadron production  at RHIC and the LHC energy at various kinematics obtained by solutions of the rcBK evolution equation (\ref{bk1}) with a initial saturation scale for proton $Q_{0p}^2=0.168\,\text{GeV}^2$ and for a nucleus within $Q_{0A}^2=3 \div 4~Q_{0p}^2$  (corresponding to the band). It is seen that the  away-side correlation has a double or single peak structure depending on the definition of the trigger (or the leading particle) and kinematics. Namely, if the leading particle is selected a prompt photon with $p^\gamma_T\ge p^h_T$, then the corresponding coincidence probability $CP_{h}(\Delta \phi)$ defined via \eq{cp1}, can have a double peak structure at $ \Delta\phi=\pi$. But if the leading particle is selected to be a hadron with $p^h_T> p^\gamma_T$, then the  coincidence probability $CP_{\gamma}(\Delta \phi)$ defined via \eq{cp2}, has a single peak structure at $\Delta\phi=\pi$. In order to understand this phenomenon, first note that the cross section of semi inclusive photon-hadron production in quark-nucleus collisions given by \eq{cs}, becomes zero for:
\begin{equation} \label{ck}
 p_T=|\vec{l_T} + \vec{p_T}^\gamma|=0.   
\end{equation}
This is simply because if the projectile parton is already without any transverse momentum, the production rate of photon-hadron should go to zero and off-shell photon remains as part of  projectile hadron wavefunction. In other words, in order the higher Fock components of projectile hadron wavefunction to be resolved and a photon to be radiated,  the projectile quark should interact with small-x target via exchanging transverse momentum. The necessary kinematics for having a local minimum for the cross-section of photon-hadron production can be readily obtained from \eq{ck} by using relations given in \eq{qh-k}, namely $l_T=p_T^h/z_f$ and the fact that for the fragmentation fraction we have $z_f^{min}\leq z_f\leq 1$. Therefore we obtain, 
\begin{eqnarray} 
z_T=\frac{p_T^h}{p_T^\gamma} &\le& 1 , \label{c-1} \\
p_T^\gamma\frac{(e^{\eta_h}+e^{\eta_\gamma})}{\sqrt{S}}&\le&1. \label{c-2}\
\end{eqnarray} 
Note that in our approach, the projectile is treated in the collinear factorization \cite{pho-cgc}. Therefore, radiation of photon from quark at this level has the standard features of  pQCD, including the back-to-back correlation in the transverse momentum. Moreover, due to multiple scatterings with target, the cross-section of photon-hadron production should have a local minimum for the back-to-back production provided the kinematics conditions given in Eqs.\,(\ref{c-1},\ref{c-2}) are satisfied.  However, because of convolution with fragmentation and parton distribution functions, the local minimum will not be zero but gets smeared out.  On the other hand, the product of $p_T^2N_F(p_T,x_g)$ in \eq{cs} has a maximum  when the transverse momentum $p_T$ approaches the saturation scale. As a result, a double peak structure appears for the away-side correlation.  Note that the integrand in the coincidence probability samples smaller transverse momentum for photon than hadron for the same reason we already mentioned, namely a photon can be produced if the parton already acquired a transverse momentum impulse. In the case that the trigger particle is selected to be a hadron rather than a prompt-photon, one should perform the integral over the transverse momentum of the hadron which is larger than transverse momentum of photon $z_T>1$, violating the condition given in \eq{c-1}, and  consequently the local minimum at $\Delta \phi=\pi$ is washed away and as a result the double peak structure will be fused to a single peak.  This can be clearly seen in Figs.\,(\ref{fig1},\ref{fig2}) at both RHIC and the LHC.

In order to further investigate the  consequences of the conditions given in Eqs\,(\ref{c-1},\ref{c-2}), let us defined the azimuthal correlation in the following form \cite{ja}, 
\begin{equation}\label{az}
P(\Delta \phi)={d\sigma^{p\, A \rightarrow h(p_T^h)\,\gamma(p_T^\gamma)\, X}
\over d^2\vec{b_t}\,p_T^h dp_T^h\, p_T^\gamma dp_T^\gamma\, d\eta_{\gamma}\, d\eta_h\, d\phi} [\Delta \phi]/ {d\sigma^{p\, A \rightarrow h(p_T^h)\,\gamma(p_T^\gamma)\, X}
\over d^2\vec{b_t}\,p_T^h dp_T^h\, p_T^\gamma dp_T^\gamma\, d\eta_{\gamma}\, d\eta_h\, d\phi} [\Delta \phi= \Delta \phi_c],
\end{equation}
which has the meaning of the probability of the semi-inclusive photon-hadron pair production at a certain kinematics and angle  $\Delta \phi$, triggering the same production with the same kinematics at  a fixed reference angle $\Delta \phi_c=\pi/2$. The correlation defined in \eq{az} may be more challenging to measure compared to the coincidence probability defined in Eq.\,(\ref{cp1},\ref{cp2}),  due to the so-called underlying event dependence, but it its free from the extra integrals over transverse momenta and  this facilitates to clearly  examine the conditions in Eqs.\,(\ref{c-1},\ref{c-2}). In a sense the correlation defined in  \eq{az}  can be considered as a snap shot of the integrand in the coincidence probability defined in Eq.\,(\ref{cp1},\ref{cp2}).

In \fig{fig2-pt}, we show the photon-hadron correlation $P(\Delta \phi)$ defined in \eq{az} at forward rapidity $\eta_h=\eta_{\gamma}=3$ for various transverse momenta  of produced prompt photon $p_T^\gamma$, and hadron $p_T^h$ at RHIC and the LHC for minimum bias pA collisions. The initial saturation scale for proton $Q^2_{0p}=0.168\,\text{GeV}^2$ and nuclei $Q_{0A}^2=3 Q_{0p}^2$ are fixed for all curves. It is clearly seen that the photon-hadron away-side correlations can have a double-peak structure both at RHIC and the LHC for the kinematics satisfying the conditions in Eqs.\,(\ref{c-1},\ref{c-2}), and the away-side double-peak correlations will evolve to a single peak  structure for kinematics outside of region defined by Eqs.\,(\ref{c-1},\ref{c-2}).  
\begin{figure}[t]       
                               \includegraphics[width=7.5 cm] {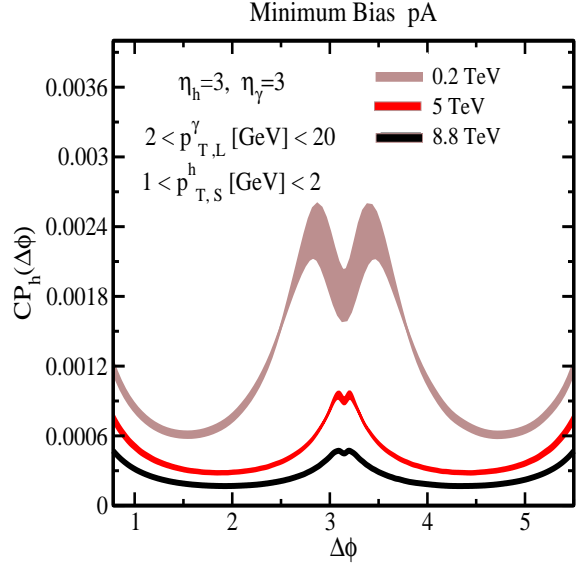}                 
\caption{ The $\gamma-\pi^0$ azimuthal correlation $CP_{h}(\Delta \phi)$ defined in  \eq{cp1} in minimum-bias pA collisions at forward rapidity $\eta_h=\eta_\gamma=3$ for various energies at RHIC and the LHC. }
\label{fig2-all}
\end{figure}

In high-energy collisions, the produced parton on average have intrinsic transverse momentum of order of the saturation scale.  By increasing the energy or density or decreasing the transverse momentum of the probe, the saturation scale $Q_s$ increases and consequently  this washes away  the intrinsic back-to-back correlations and the away-side correlation is suppressed. Numerically, a bigger saturation scale, pushes the unintegrated gluon density profile to larger transverse momentum. As a result, the single inclusive production (either hadron or prompt photon) cross section (the denumenator in the coincidence probability) is enhanced, while the two-particle correlated cross section \eq{cs} is suppressed by a larger saturation scale $Q_s$.  Therefore, the coincidence  probability defined in Eqs.\,(\ref{cp1},\ref{cp2}), decreases with increasing the saturation scale and  
we expect  that the photon-hadron away-side correlation at the LHC to be smaller than  RHIC (at the same rapidity and transverse momenta of associated and leading particle). This can be clearly seen in \fig{fig2-all} where we compare the coincidence probability $CP_{h}(\Delta \phi)$ obtained at various energies.  Moreover, the saturation scale grows with density, therefore the away-side correlations in pA collisions should be more suppressed compared to pp collisions at the same kinematics, see Figs.\,(\ref{fig1},\ref{fig2}). 

It is seen from Figs.\,(\ref{fig1},\ref{fig2},\ref{fig2-pt}) that generally at a fixed rapidity and energy, the suppression of away-side $\gamma-h$ correlation is larger for a case that $z_T>1$. This effect can be traced back to the fact that $\gamma-h$ pairs with $z_T>1$ probe lower $x_g$-region compared to the cases that $z_T<1$. This can be understood by rewriting the definition of $x_g$ in \eq{qh-k} which appears in the unintegrated gluon density  in term of $x_T$, namely 
$x_g=\frac{p_T^\gamma}{\sqrt{S}}\left( e^{-\eta_\gamma}+ \frac{z_T}{z_f}\, e^{-\eta_{h}}\right)$. Therefore,  $\gamma-h$ pairs production with $z_T>1$ have a lower $x_g$ and consequently the suppression due to saturation will be larger.

In \fig{fig2}, we show $CP_{h}(\Delta \phi)$ and $CP_{\gamma}(\Delta \phi)$ at the LHC $\sqrt{S}=8.8$ TeV at forward rapidity $\eta_h=\eta_{\gamma}=3$ in minimum-bias pA and pp collisions for two bins of transverse momenta of associated prompt photon $p^{\gamma}_{T,S}$ and hadron $p^{h}_{T,S}$, and the corresponding leading hadron $p^{\gamma}_{T,L}$ and prompt photon $p^{\gamma}_{T,L}$. Namely 
in top and lower panel  we performed the integral for the associated particle within $[1,2]$ GeV and $[1, 6]$ GeV (and for the corresponding leading particle within $[2,20]$ GeV and $[6, 20]$ GeV), respectively. 
The correlation signal enhances by  increasing the transverse momenta interval of associated particle. This is simply because in Eqs.\,(\ref{cp1},\ref{cp2}), by construction, the integrals over the leading particle is mainly canceled out in the ratio, and the correlation becomes proportional to the integral  over the associated particle. For higher transverse momenta bins, the saturation scale is smaller and the back-to-back correlation is  restored.  Notice that since the rcBK evolution solution is not reliable at high transverse momentum we had to impose upper limit cut for the integrals over transverse momenta of the leading particle in  Eq.\,(\ref{cp1},\ref{cp2}).  However, the cross-sections drop so fast with transverse momentum at forward rapidities that this upper cutoff should not make a big difference.

The azimuthal correlation $CP_{h}(\Delta \phi)$ defined in \eq{cp1} is generally bigger than the corresponding correlation $CP_{\gamma}(\Delta \phi)$  defined in \eq{cp2}, at the same kinematics. This is because when the trigger particle is taken a prompt photon, in \eq{cp1}, the electromagnetic coupling $\alpha_{em}$ drops out in the ratio of two cross sections and that enhances the signal compared to the case that the trigger particle is selected to be a hadron.  This can be seen in Figs.\,(\ref{fig1},\ref{fig2}).
\begin{figure}[t]                                                            
                           \includegraphics[width=7.5 cm] {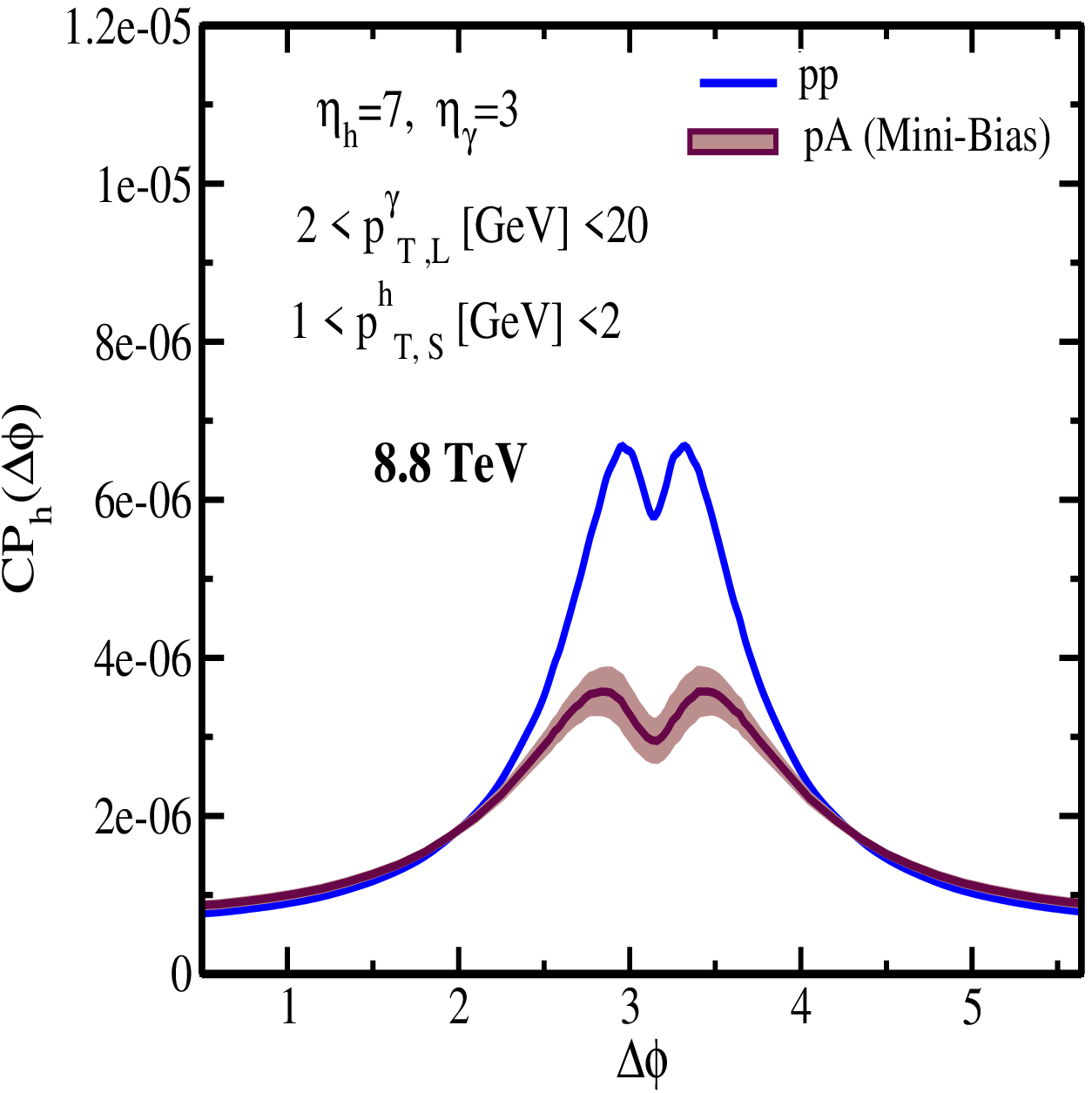}
                             \includegraphics[width=7.5 cm] {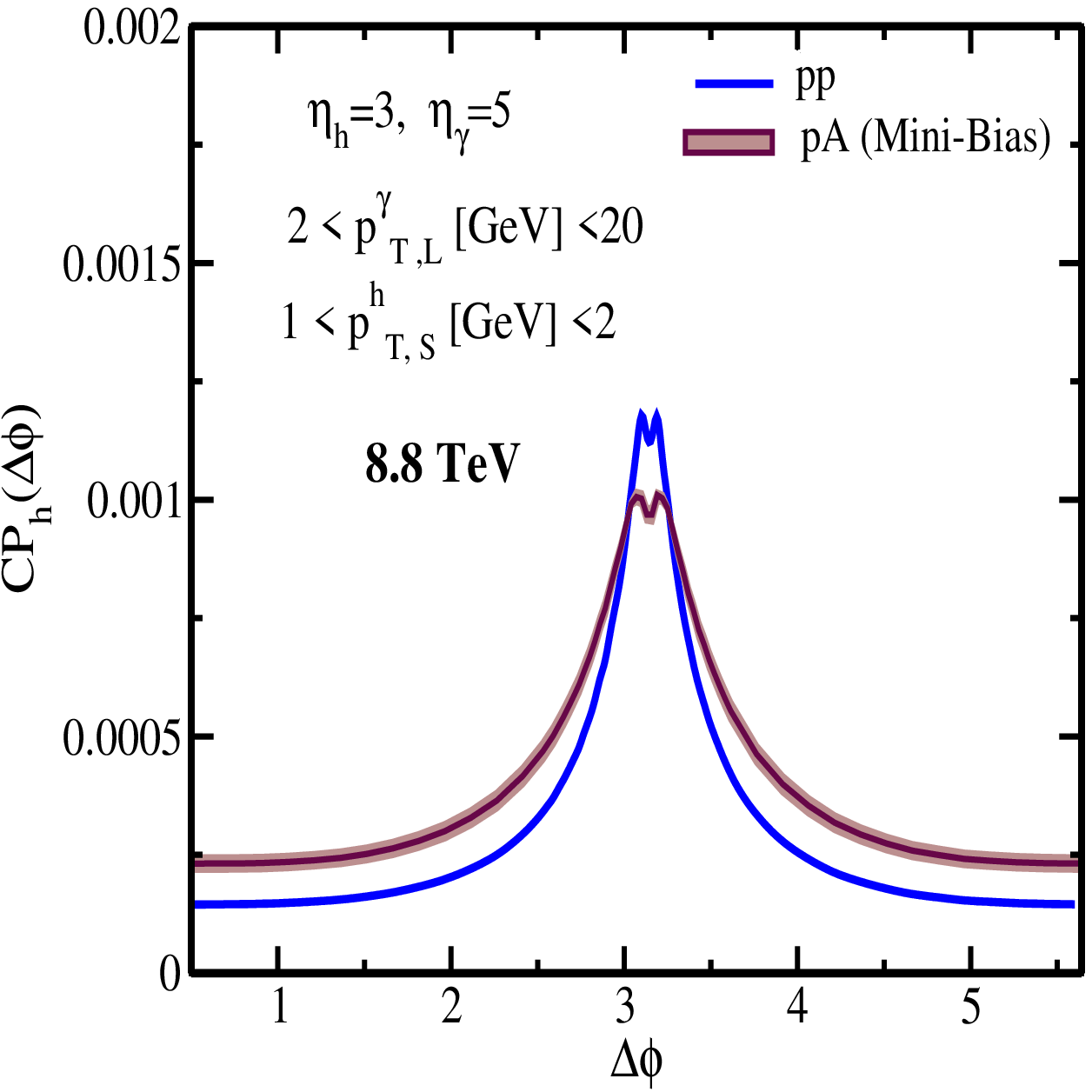}     

\caption{ 
The $\gamma-\pi^0$ azimuthal correlation $CP_{h}(\Delta \phi)$  defined in \eq{cp1} in minimum-bias pA and pp collisions at the LHC $\sqrt{S}=8.8$ TeV at different rapidities of the produced hadron  $\eta_h$ and prompt photon $\eta_\gamma$.}
\label{fig3}
\end{figure}

As we already pointed out, the double peak structure for the photon-hadron coincidence probability $CP_{h}(\Delta \phi)$  at $\Delta \phi \approx \pi$ is due to the interplay between a local minimum for the cross-section at $p_T\approx 0$ and two maxima for the cross-section when $p_T\approx Q_s$. The double-peak structure can be stretched  out and becomes more pronounced by measuring the associated hadron at about or higher rapidity than the  trigger prompt photon, i.e. $\eta_h\ge\eta_\gamma$.  This is due to the fact that because of kinematic limit for more forward production, the integrand of the associated hadron in $CP_{h}(\Delta \phi)$  is relatively shifted to lower transverse momentum and consequently the conditions for local minimum in Eqs.\,(\ref{c-1},\ref{c-2})  are satisfied  while at the same time, the saturation scale increases for more forward production leading to an enhancement of the two local maxima. In Figs.\,(\ref{fig1},\ref{fig3}), we show this effect by comparing the azimuthal correlations at different rapidities  $\eta_h$ and $\eta_\gamma$ at RHIC and the LHC.

\begin{figure}[t]       
                               \includegraphics[width=7.5 cm] {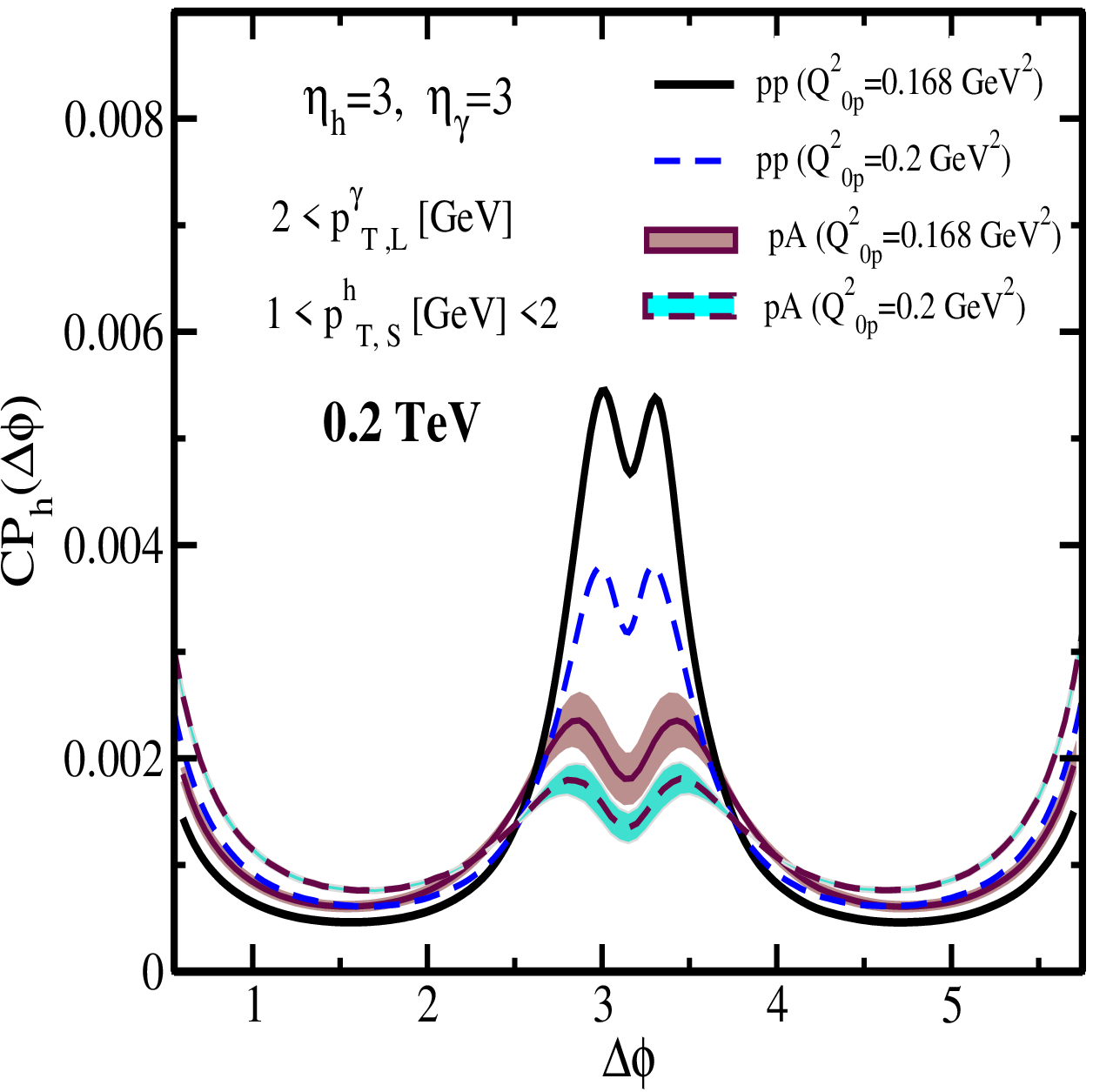}  
                               \includegraphics[width=7.5 cm] {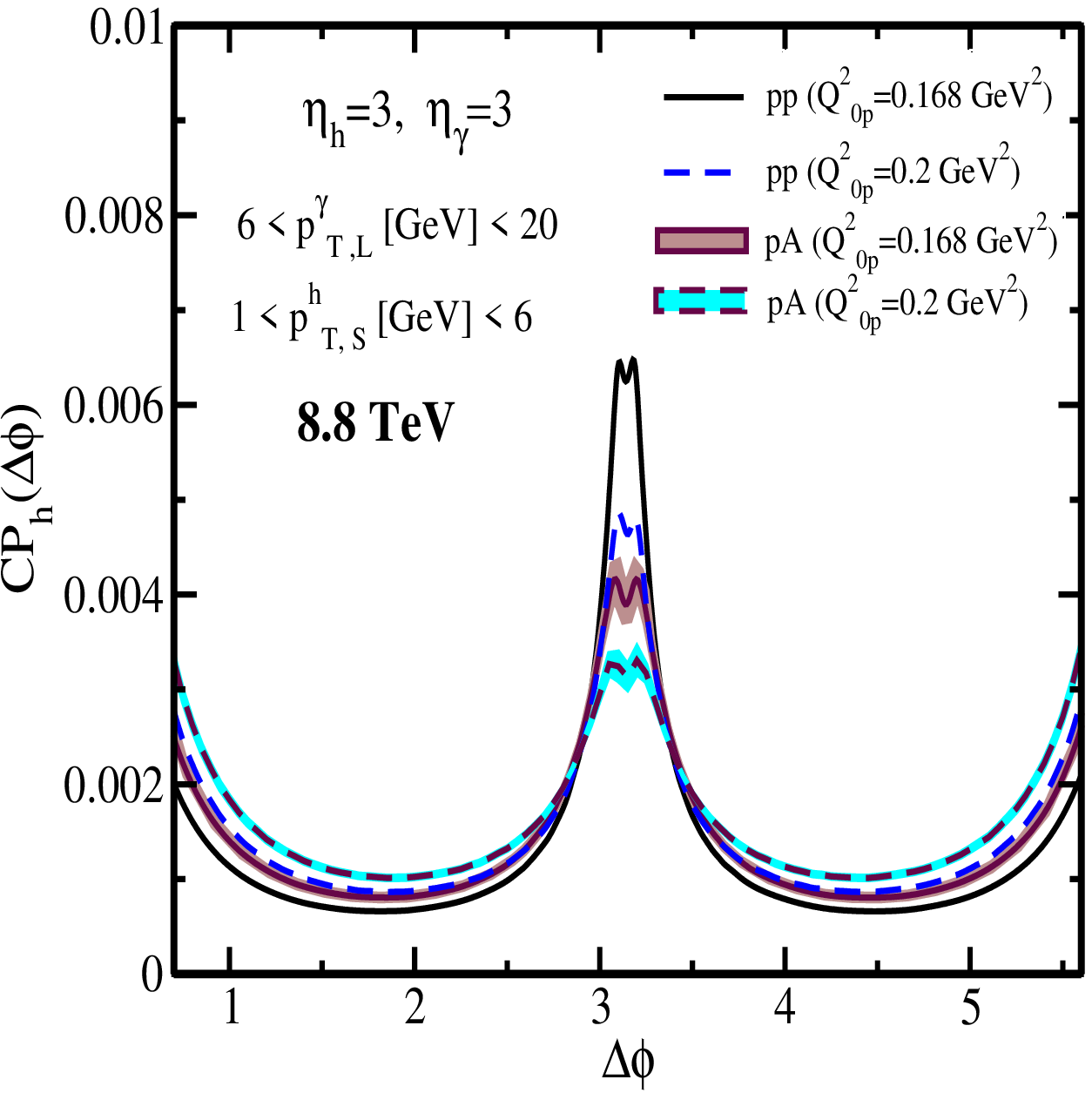}                              
\caption{ The $\gamma-\pi^0$ azimuthal correlation $CP_{h}(\Delta \phi)$ in minimum-bias pA and pp collisions at forward rapidity at RHIC and the LHC. The  curves are obtained by the rcBK equation with two different initial saturation scale of proton $Q_{0p}^2=0.168$ and $0.2\,\text{GeV}^2$ and  the corresponding initial saturation scale of the nucleus within $Q_{0A}^2=3\div 4 Q_{0p}^2$.   }
\label{fig2-u}
\end{figure}

Although the main features of the photon-hadron correlations, e.g. the double or single peak structure and  decorrelation with energy/rapidity, density and transverse momentum seem to be robust and understandable due to the non-linear gluon saturation dynamics,  there is some uncertainties on the magnitude of the correlation obtained in our approach. These uncertainties are due to the fact that with available worldwide small-x experimental data it is not yet possible to uniquely fix the parameters of the rcBK evolution equation and the initial saturation scale of proton and nucleus \cite{j1,jav1,me-jamal1,jm,raj}. To highlight our main uncertainties, in \fig{fig2-u} we show  $CP_{h}(\Delta \phi)$ for minimum-bias pA and pp collisions at forward rapidity at RHIC and the LHC, with two different initial saturation scale of proton, namely $Q_{0p}^2=0.168$ and $0.2\,\text{GeV}^2$ corresponding to $\gamma=1.19$ and $\gamma=1$ in \eq{mv} respectively  \cite{jav1}, and the initial saturation scale of nuclei (gold and lead) within $Q_{0A}^2=3\div 4 Q_{0p}^2$. Although both values of $Q_{0p}$ (or $\gamma$) are extracted from a fit to HERA data on the proton target at small-x \cite{jav1} (with a good $\chi^2$), the recent LHC data seems to favor the parameter set with $\gamma>1$ (or lower value for $Q_{0p}$) \cite{j1}. The uncertainties in the initial scale for proton will bring even larger uncertainties in determining the parameters of the rcBK equation for the case of nuclear target\footnote{This is partly due to the fact that  solution of the rcBK equation in the presence of impact-parameter is not yet available.}.  Therefore, the upcoming LHC data on pA collisions can provide crucial complementary constrain on the rcBK evolution equation and small-x physics.  For other measurements sensitive to the saturation physics at the LHC, see Refs.\,\cite{ja,me-pa,me-jamal1}.

Note that the semi-inclusive photon-hadron cross-section in \eq{cs}  has collinear singularity. Therefore, one should first treat the collinear singularity for the near-side jet $\Delta \phi\approx 0$  in a same fashion as was done for the inclusive prompt photon production in \eq{pho2} by introducing the quark-photon fragmentation function. Therefore, our results at near-side  $\Delta \phi \approx 0$ should be less reliable. However, one should bear in mind that the integrand in the azimuthal correlation generally samples lower transverse momenta for the away-side correlations $\Delta \phi \approx \pi$ than for near-side ones $\Delta \phi \approx 0$. Therefore, 
here we only focused on the away-side correlations which is a sensitive probe of small-x physics and gluon saturation.

\begin{figure}[t]                                                            
                                  \includegraphics[width=7.5 cm] {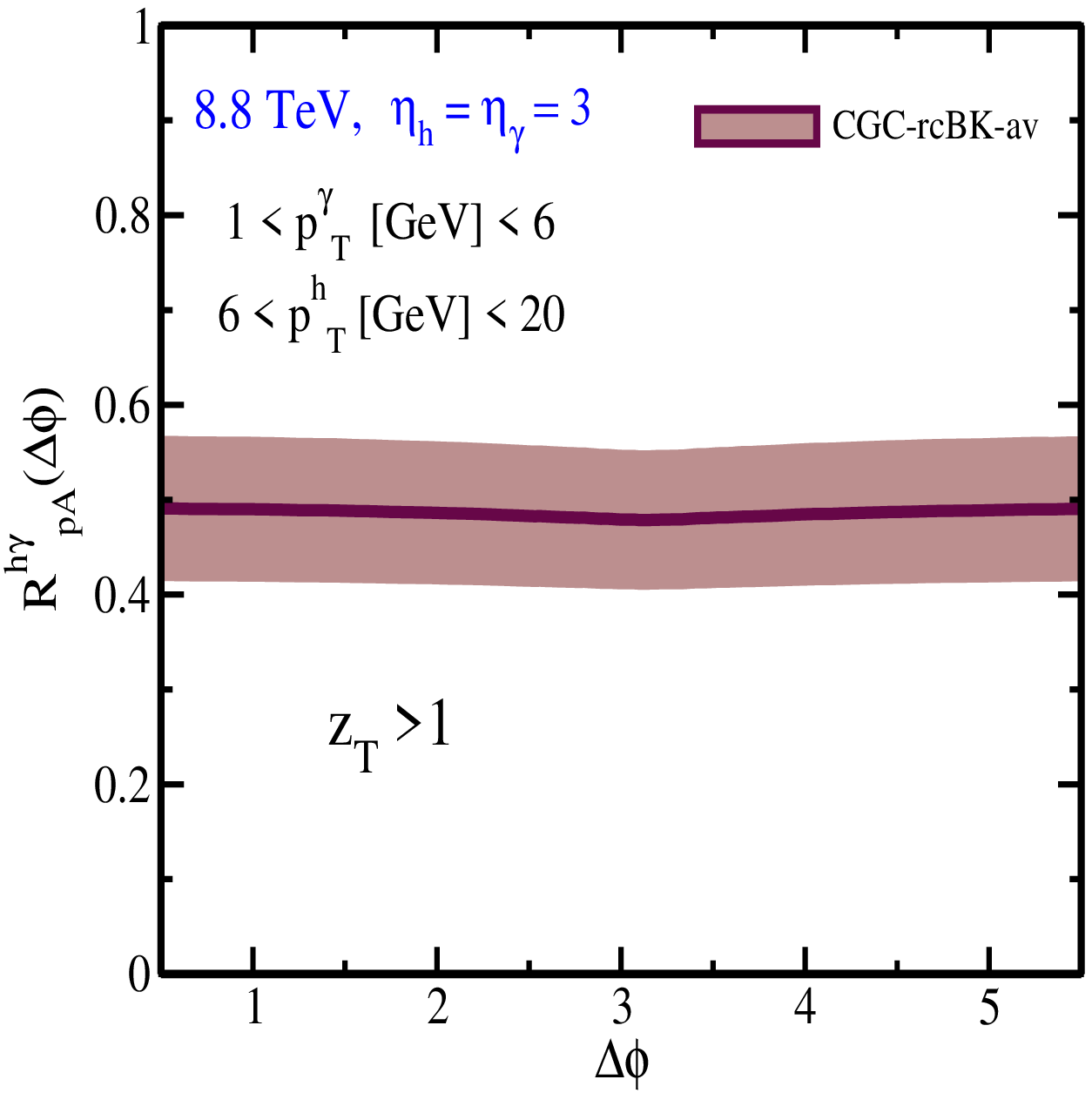}   
                                  \includegraphics[width=7.5 cm] {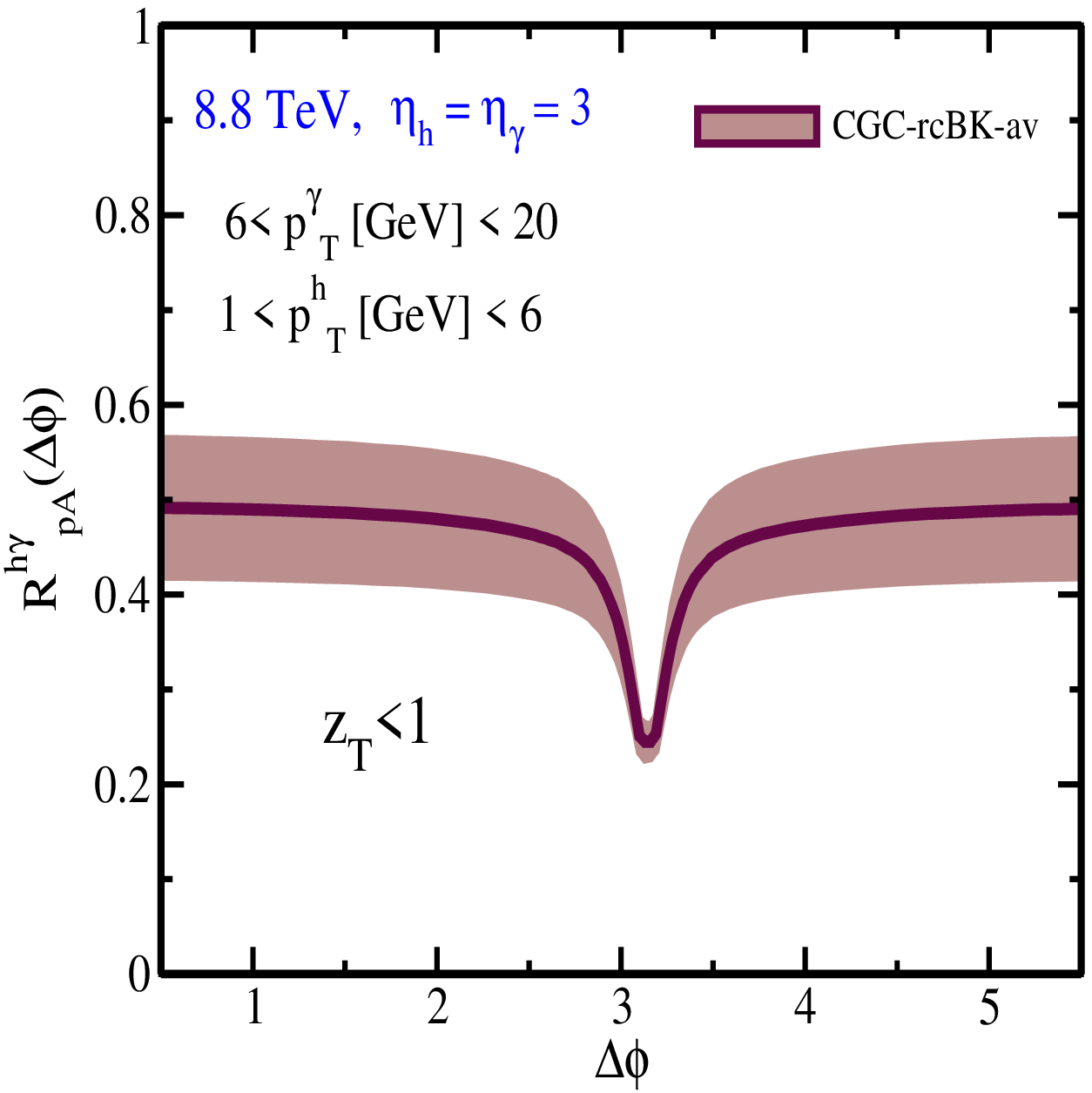}       
\caption{The nuclear modification factor $R_{pA}^{h\gamma }$ for semi-inclusive photon-hadron ($\gamma-\pi^0$) production defined in \eq{rp1}  as a function of $\Delta \phi$ at the LHC in minimum-bias pA collisions at forward rapidity $\eta_h=\eta_\gamma=3$ for two different bins of transverse momenta  of produced prompt photon $p^\gamma_T$ and hadron $p^h_T$, namely $z_T<1$ (right) and $z_T>1$ (left). The band (CGC-rcBK-av) incorporates the uncertainties due to  variation  of the initial saturation scale in the rcBK evolution equation. }
\label{fig-r1}
\end{figure}

\begin{figure}[t]                                                            
                                  \includegraphics[width=11 cm] {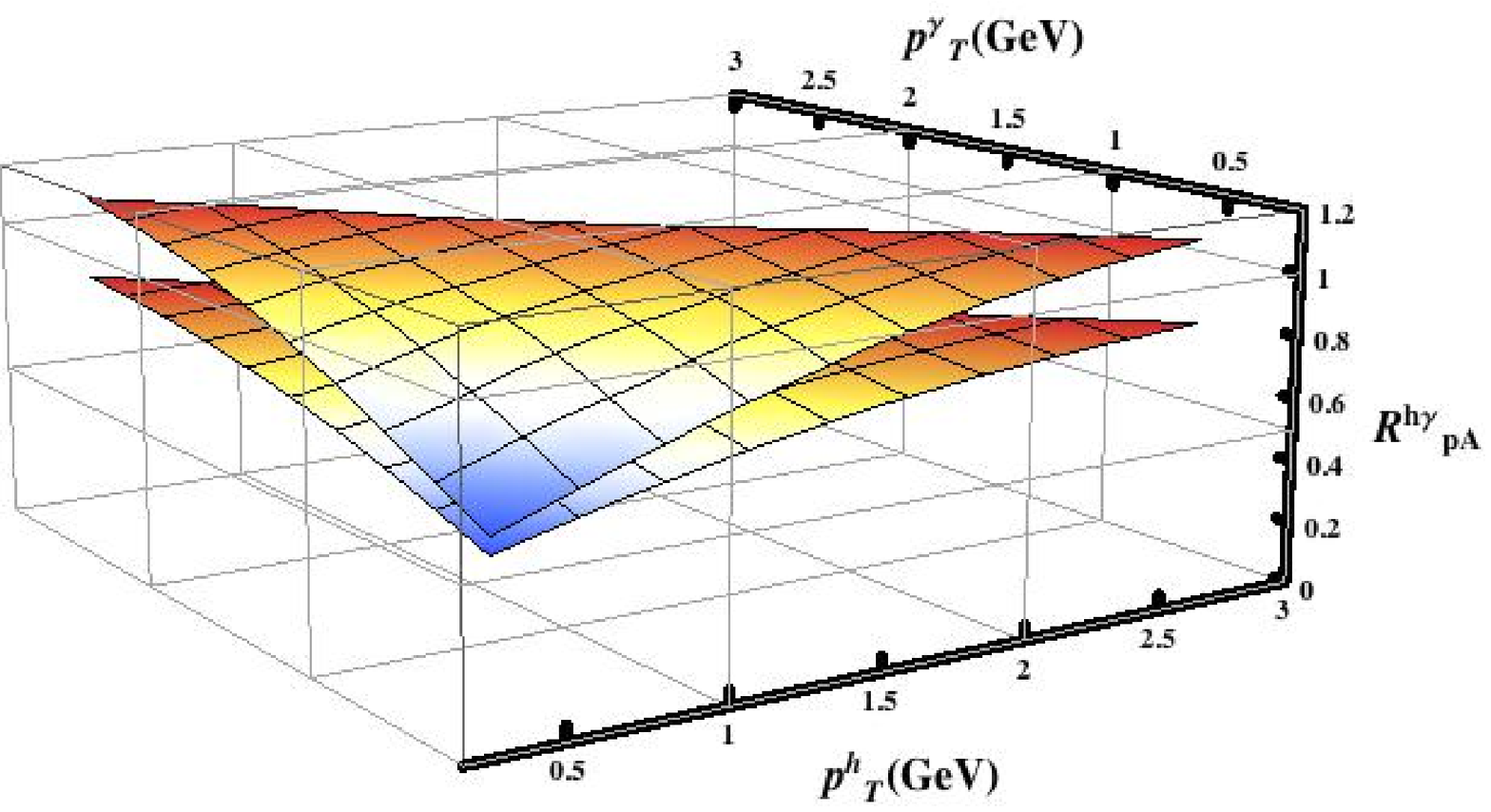}     
                                   \includegraphics[width=11 cm] {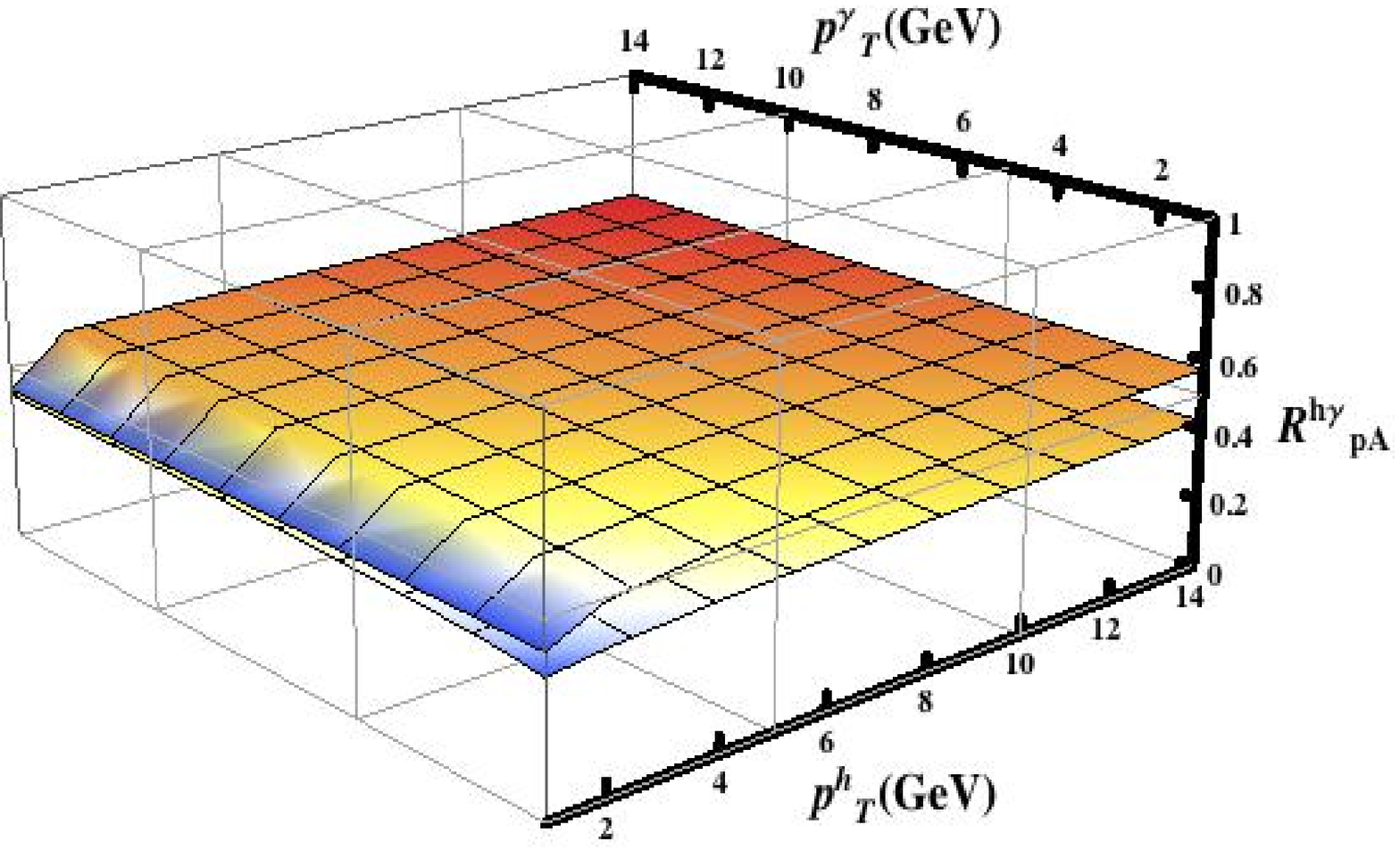} 
\caption{The nuclear modification factor $R_{pA}^{h\gamma }$ for semi-inclusive $\gamma-\pi^0$ production defined in \eq{rp2}
as a function of transverse momentum of produced prompt photon $p^\gamma_T$ and hadron $p^h_T$ in minimum-bias pA collisions at RHIC $\sqrt{S}=0.2$ TeV  at $\eta_h=\eta_\gamma=4$ (top panel) and  the LHC $\sqrt{S}=8.8$ TeV  at $\eta_h=\eta_\gamma=3$ (lower panel). Two surfaces are obtained from the solutions of the rcBK  evolution equation with two different initial saturation scale of the nucleus, see the text for the details. }
\label{fig-rl}
\end{figure}

In \fig{fig-r1}, we show the nuclear modification factor $R_{pA}^{h\gamma }$ for semi-inclusive photon-hadron pair production defined in \eq{rp1}  as a function of $\Delta \phi$ at the LHC energy $\sqrt{S}=8.8$ TeV and $\eta_h=\eta_{\gamma}=3$ for two different transverse momenta bins of produced prompt photon $p^\gamma_T$ and hadron $p^h_T$ (the integral is performed over the given interval of transverse momenta).  Similar to previous plots, the band  (CGC-rcBK-av) in \fig{fig-r1} comes from the rcBK solutions incorporating the uncertainties associated to  a variation of the initial saturation scale of the nucleus in a range consistent with previous studies of DIS structure functions as well as particle  production in minimum-bias pp, pA and AA collisions in the CGC formalism. One may therefore expect that the possible effects of fluctuations on particle production is effectively contained in our error band. The away-side nuclear modification $R_{pA}^{h\gamma }$ at $\Delta\phi\approx \pi$ is dramatically suppressed with a lower peak structure when  the transverse momentum bin of the produced  prompt photon is larger than hadron $z_T<1$. This is fully in accordance with the photon-hadron decorrelation in pA compared to pp collisions, and  conditions given in   Eqs.\,(\ref{c-1},\ref{c-2})  for the existence of the local minimum for the away-side photon-hadron production.  Note that the sensitivity to the transverse momenta or the ratio $z_T$ only manifests itself at around  $\Delta\phi\approx \pi$. 

Finally, in \fig{fig-rl}, we show the two-dimensional nuclear modification factor $R_{pA}^{h\gamma }$ for semi-inclusive photon-hadron production defined in \eq{rp2}  as a function of transverse momentum  of produced prompt photon $p^\gamma_T$ and hadron $p^h_T$  at RHIC $\sqrt{S}=0.2$ TeV at $\eta_h=\eta_{\gamma}=4$ (top panel) and at the LHC $\sqrt{S}=8.8$ TeV at $\eta_h=\eta_{\gamma}=3$ (lower panel). 
 The area between two surfaces in \fig{fig-rl}, similar to \fig{fig-r1} (the band labeled by CGC-rcBK-av) shows the uncertainties associated to  the variation of the initial saturation scale of the nucleus. It is seen that at the LHC energy $\sqrt{S}=8.8$ TeV, the nuclear modification factor $R_{pA}^{h\gamma }$ is more suppressed compared to RHIC and also is more flat. We recall that the semi-inclusive photon-hadron cross-section \eq{cs}  is not equal to the product of cross-sections of single inclusive prompt photon and hadron production given in \eq{pho2} and \eq{final}.   Note that in Ref.\,\cite{ja} it was shown that at RHIC for the single inclusive prompt photon production, a good portion of the suppression at forward rapidities is due to the projectile being a deuteron rather than a proton. Here for a comparison with  pA run at the LHC energy and in order to discard possible suppression associated to isospin effect \cite{ja}, we have only considered proton-nucleus collisions at the RHIC energy which can be also useful for the future pA run at RHIC. We check that similar to Ref.\,\cite{ja}, discarding the fragmentation photon contribution from the cross-section, will not affect our results for $R_{pA}^{h\gamma }$  significantly. Nevertheless, a detailed study of the semi-inclusive photon-hadron production in the presence of  isolation cut is beyond the scope of the current paper.

%---------------------------------------------------------------------
\section{Summary} 
We have investigated semi-inclusive prompt photon-hadron production in high-energy pp and pA collisions within the CGC framework by using the running-coupling BK equation. We provided detailed predictions for the coincidence probability of photon-hadron correlations and showed that such correlations exhibit novel feature, namely the away-side correlations can have a double or single peak structure depending on the trigger particle selection and kinematics. The correlations have a double-peak structure by selecting $\gamma-h$ pairs within the kinematics region satisfying the conditions in Eqs.\,(\ref{c-1},\ref{c-2}), and the away-side double-peak correlations will evolve to a single peak  structure for kinematics outside of that region. We showed that this feature can be understood by QCD saturation dynamics. The double-peak structure for the azimuthal correlations has been also recently reported for other electromagnetic probe, namely the Drell-Yan Lepton-Pair-Jet correlation in pA collisions \cite{dy} while it is absent for dihadron production \cite{di,di-e}.  The decorrelation of the away-side photon-hadron production with energy, rapidity, density and transverse momentum of the probe  is very similar to the dihadron production in pA collisions and can be understood in the CGC framework. If experimentally confirmed, this will provide a significant evidence in favor of the universality of  particle production in the QCD saturation picture at high-energy.

In a sense, the double-peak structure for $\gamma-h$ correlations resembles the long-range azimuthal correlations for the produced charged hadron pairs, observed in high-multiplicity events in pp collisions at the LHC, the so-called ridge phenomenon \cite{ridge-cms}.  Although, the ridge is a feature on a near-side  $\Delta \phi\approx 0$ of the two particle correlations, while the $\gamma-h$ double-peak structure  is a away-side feature. In both cases, a second local maximum occurs because of angular collimation due to the presence of the saturation scale in the system, and the effect shows up within a kinematics window which is dictated by the saturation scale\footnote{The author thanks Raju Venugopalan for pointing out the possible similarity between these two phenomena.} \cite{ridge,ridge0,alex}. Similar to the ridge, the double-peak structure here can survive up to rather large rapidity (see \fig{fig3}), and in both cases, one expects that the same mechanism to be responsible for the self-deconstruction of the effect namely decorrelation  at very high-energy \cite{alex}.

We also showed  that the ratio $z_T = p_T^{h}/p_T^{\gamma}$  is a sensitive parameter to the saturation region and controls the away-side $\gamma-h$ suppression
 in high-energy pp and pA collisions.  

We studied the ratio of single inclusive prompt photon to hadron production $\gamma^{\text{inclusive}}/\pi^0$ in pp and pA collisions at RHIC and the LHC at various rapidities. We found that the ratio $\gamma^{\text{inclusive}}/\pi^0$  is very similar for high-energy pp and pA collisions at forward rapidities at high transverse momentum, and it increases with rapidity while it decreases with energy. We also provided predictions for the nuclear modification factor for the semi-inclusive photon-hadron pair production $R_{pA}^{h\gamma }$ in pA collisions at RHIC and the LHC at forward rapidities. We showed that  the two-dimensional $R_{pA}^{h\gamma }$ is generally more flat at the LHC compared to RHIC at forward rapidities. We found that the suppression of  the nuclear modification factor for semi-inclusive photon-hadron production is comparable to that for single inclusive hadron \cite{me-jamal1} and prompt photons \cite{ja} production in pA collisions at forward rapidities.

%-----------------------------------------------------------------------
\begin{acknowledgments}
The author would like to thank Thomas Peitzmann and Richard Seto for useful discussions which led to this paper. 
The author is greatful to Raju Venugopalan for a careful reading of the manuscript and useful comments. 
It is a great pleasure to thank Adrian Dumitru and Jamal Jalilian-Marian for fruitful conversations at the early stage of this work. 
This work is supported in part by Fondecyt grants 1110781.
\end{acknowledgments}

%---------------------------------------------------------------------


\begin{thebibliography}{99}

\bibitem{sg}
L. V. Gribov, E. M. Levin and M. G. Ryskin, Phys. Rept. {\bf 100}, 1 (1983); A. H. Mueller and  J-W. Qiu, Nucl. Phys. {\bf 268}, 427
(1986).

\bibitem{mv} 
  L.~D.~McLerran and R.~Venugopalan,
  %``Computing quark and gluon distribution functions for very large nuclei,''
  Phys.\ Rev.\ D {\bf 49}, 2233 (1994); Phys. Rev. {\bf D49}, {\it ibid}. {\bf 49}, 3352 (1994); {\it ibid}. {\bf 50}, 2225 (1994).
\bibitem{cgc-review1}
%\cite{Gelis:2010nm}
E. Iancu, A. Leonidov and L. McLerran, hep-ph/0202270; E. Iancu and R. Venugopalan, hep-ph/0303204; F. Gelis, E. Iancu, J. Jalilian-Marian and R. Venugopalan, Ann. Rev. Part. Nucl Sci. {\bf 60}, 463 (2010) and references therein. 
\bibitem{e-lhc}
CMS Collaboration, Phys. Rev. Lett. {\bf 105} 022002 (2010) [arXiv:1005.3299]; 
ALICE Collaboration, Phys. Rev. Lett. {\bf 105}, 252301 (2010) [arXiv:1011.3916];
CMS Collaboration, J. High Energy Phys. {\bf 08}, 141 (2011) [arXiv:1107.4800];  ATLAS Collaboration, Phys. Lett. {\bf B710}, 363 (2012) [arXiv:1108.6027].

\bibitem{m1}
E. Levin and A. H. Rezaeian, Phys. Rev. {\bf D82}, 014022 (2010); Phys. Rev. {\bf D83}, 114001 (2011). 

\bibitem{tr}
P. Tribedy and R. Venugopalan, Nucl. Phys. A850, 136-156 (2011) [Erratum-ibid. {\bf A859}, 185 (2011)];
A. Dumitru and Y. Nara, Phys. Rev. {\bf C85}, 034907 (2012).
\bibitem{tr1}
D. Kharzeev, E. Levin and M. Nardi, Nucl. Phys. {\bf A747}, 609 (2005). 
\bibitem{j1}
J. L. Albacete and A. Dumitru,  arXiv:1011.5161. 
\bibitem{me-pa}
A. H. Rezaeian, Phys.Rev. {\bf D85}, 014028 (2012) [arXiv:1111.2312]; arXiv:1208.0026; arXiv:1110.6642; E. Levin and A. H. Rezaeian, Phys. Rev. {\bf D82}, 054003 (2010),  arXiv:1011.3591.


\bibitem{ridge}
A. Dumitru, K. Dusling, F. Gelis, J. Jalilian-Marian, T. Lappi and R. Venugopalan, Phys. Lett. {\bf B697}, 21 (2011); 
K. Dusling and R. Venugopalan, arXiv:1201.2658. 

\bibitem{ridge0}
E. Levin and A. H. Rezaeian, Phys.Rev. {\bf D84}, 034031 (2011).

\bibitem{ja}
 J. Jalilian-Marian and A. H. Rezaeian, Phys. Rev. {\bf D86}, 034016 (2012).

\bibitem{di} 
C. Marquet, Nucl. Phys. {\bf A796}, 41 (2007); K. Tuchin, Nucl. Phys. {\bf A846}, 83 (2010);
  J.~L.~Albacete and C.~Marquet,
  %``Azimuthal correlations of forward di-hadrons in d+Au collisions at RHIC in the Color Glass Condensate,''
  Phys.\ Rev.\ Lett.\  {\bf 105}, 162301 (2010); F. Dominguez, B. W. Xiao and F. Yuan, Phys. Rev. Lett. 106, 022301 (2011); F. Dominguez, et al., Phys. Rev. {\bf D83}, 105005 (2011);  A. Stasto, Bo-Wen Xiao and F. Yuan, arXiv:1109.1817; K. Kutak and S. Sapeta, arXiv:1205.5035; T. Lappi and H. M\"antysaari, arXiv:1207.6920. 
  


\bibitem{pho-cgc} 
  F.~Gelis and J.~Jalilian-Marian,
  %``Photon production in high-energy proton nucleus collisions,''
  Phys.\ Rev.\ D {\bf 66}, 014021 (2002); 
%\cite{Baier:2004tj}
  R.~Baier, A.~H.~Mueller and D.~Schiff,
  %``Saturation and shadowing in high-energy proton nucleus dilepton production,''
  Nucl.\ Phys.\ A {\bf 741}, 358 (2004).
  %%CITATION = HEP-PH/0403201;%%


\bibitem{bk}
I.~Balitsky,
%``Operator expansion for high-energy scattering,''
Nucl.\ Phys.\  {\bf B463}, 99 (1996);
Y.~V.~Kovchegov,
%``Small-x F2 structure function of a nucleus including multiple pomeron
%exchanges,''
Phys.\ Rev.\   {\bf D60}, 034008 (1999);
%%CITATION = PHRVA,D60,034008;%%
%``Unitarization of the BFKL pomeron on a nucleus,''
Phys.\ Rev.\   {\bf D61}, 074018 (2000).
%%CITATION = PHRVA,D61,074018;%%
\bibitem{bb}
I. I. Balitsky, Phys. Rev. {\bf D75}, 014001 (2007) [hep-ph/0609105].
\bibitem{nlo}
I. Balitsky, G. A. Chirilli, Phys. Rev. {\bf D77}, 014019 (2008);
E. Gardi, J. Kuokkanen, K. Rummukainen and H. Weigert, Nucl. Phys. {\bf A784}, 282 (2007);
Y. V. Kovchegov and H. Weigert, Nucl. Phys. {\bf A784}, 188 (2007).
%\cite{Avsar:2011ds}
  E.~Avsar, A.~M.~Stasto, D.~N.~Triantafyllopoulos and D.~Zaslavsky,
  %``Next-to-leading and resummed BFKL evolution with saturation boundary,''
  JHEP {\bf 1110}, 138 (2011).



\bibitem{rcbk}
  J.~L.~Albacete and Y.~V.~Kovchegov,
  %``Solving high energy evolution equation including running coupling corrections,''
  Phys.\ Rev.\  {\bf D75}, 125021 (2007).
\bibitem{jav1}
J. L. Albacete, N. Armesto, J.G. Milhano, P. Quiroga Arias and C. A. Salgado, Eur. Phys.  J. {\bf C71}, 1705 (2011). 

\bibitem{m-b}
B. Z. Kopeliovich and A. H. Rezaeian, Int. J. Mod. Phys. {\bf E18}, 1629 (2009) [arXiv:0811.2024]. 

\bibitem{ipsat}
K. Golec-Biernat and M. Wusthoff, Phys. Rev. {\bf D59}, 014017 (1998) 
H. Kowalski and D. Teaney, Phys.Rev. {\bf D68}, 114005 (2003);
E. Iancu, K. Itakura and S. Munier, Phys. Lett. {\bf B590} ,199 (2004 );
H. Kowalski, L. Motyka and G. Watt, Phys. Rev. {\bf D74}, 074016 (2006); G. Watt and  H. Kowalski, Phys. Rev. {\bf D78}, 014016 (2008);
E. Gotsman, E. Levin, M. Lublinsky and U. Maor, Eur. Phys. J. {\bf C27}, 411 (2003); M. Lublinsky, Eur. Phys. J. {\bf C21}, 513 (2001);
A.H. Mueller, D.N. Triantafyllopoulos, Nucl. Phys. {\bf B640}, 331 (2002); D. N. Triantafyllopoulos, Nucl. Phys. {\bf B648}, 293 (2003); 
C. Marquet and G. Soyez, Nucl. Phys. {\bf A760}, 208 (2005).
\bibitem{novel}
For example: V. Khachatryan {\it et al.} [CMS Collaboration], Phys. Rev. Lett. {\bf 106}, 122003 (2011); JHEP {\bf 1009}, 091 (2010); J. Adams {\it et al.} [STAR Collaboration], Phys. Rev. Lett. {\bf 91}, 072304 (2003); G. Aad {\it et al.} [ATLAS Collaboration], Phys. Rev. Lett. {\bf 105}, 252303 (2010); S. Chatrchyan {\it et al.} [CMS Collaboration], Phys. Rev. {\bf C84}, 024906 (2011); K. Aamodt {\it et al.} [ALICE Collaboration], Phys. Rev. Lett. {\bf 108} 092301 (2012). 


\bibitem{di-e}
A. Adare {\it et al.} [PHENIX Collaboration], Phys. Rev. Lett. {\bf 107}, 172301 (2011);
E. Braidot, for the STAR Collaboration, Nucl. Phys. {\bf A854}, 168 (2011); E. Braidot, Ph.D. thesis,  arXiv:1102.0931. 

\bibitem{ph-ex}
 A. Adare  {\it et al.} [PHENIX Collaboration], Phys. Rev. {\bf C80}, 024908 (2009). 
\bibitem{ph-th}
X.-N. Wang and Z. Huang, Phys. Rev. {\bf C55}, 3047 (1997); X.-N. Wang, Z. Huang, and I. Sarcevic, Phys. Rev. Lett.
{\bf 77}, 231 (1996); H. Zhang, J. F. Owens, E. Wang and X.-N. Wang, Phys. Rev. Lett. {\bf 103}, 032302 (2009);
G.-Y. Qin, J. Ruppert, C. Gale, S. Jeon and G. D. Moore, Phys. Rev. {\bf C80}, 054909 (2009). 

\bibitem{jimwlk}
J. Jalilian-Marian, A. Kovner, A. Leonidov and H. Weigert, Nucl. Phys. {\bf B504}, 415 (1997); {\it ibid.}, Phys. Rev. {\bf D59}, 014014
(1999); E. Iancu, A. Leonidov and L. D. McLerran, Nucl. Phys. {\bf A692}, 583 (2001); E. Ferreiro, E. Iancu, A. Leonidov and L. D. McLerran, Nucl. Phys. {\bf A703}, 489 (2002). 
\bibitem{jimwlk1}
A. Dumitru, J. Jalilian-Marian, T. Lappi, B. Schenke and R. Venugopalan, Phys. Lett. {\bf B706}, 219 (2011); T. Lappi, Phys. Lett. {\bf B703}, 325 (2011). 

\bibitem{own}
J. F. Owens, Rev. Mod. Phys. {\bf 59}, 465 (1987).


\bibitem{me2-pho}
A. H. Rezaeian and A. Schaefer, Phys. Rev. {\bf D81}, 114032  (2010) [arXiv:0908.3695]; B. Z. Kopeliovich, A. H. Rezaeian, H. J. Pirner and I. Schmidt, Phys. Lett. {\bf B653}, 210 (2007) [arXiv:0704.0642]; B. Z. Kopeliovich, E. Levin, A. H. Rezaeian and I. Schmidt, Phys. Lett. {\bf B675}, 190 (2009); B. Z. Kopeliovich, H. J. Pirner, A.H. Rezaeian, I. Schmidt,
Phys. Rev. {\bf D77}, 034011 (2008); M. V. T. Machado and C. B. Mariotto, Eur. Phys. J. {\bf C61}, 871 (2009). 



\bibitem{dhj}
  A.~Dumitru, A.~Hayashigaki and  J.~Jalilian-Marian,
  %``The Color glass condensate and hadron production in the forward region,''
  Nucl.\ Phys.\  {\bf A765}, 464 (2006). 
  
\bibitem{inel}
T.~Altinoluk and A.~Kovner,
  %``Particle Production at High Energy and Large Transverse Momentum - 'The Hybrid Formalism' Revisited,''
  Phys.\ Rev.\  {\bf D83}, 105004 (2011).
\bibitem{me-jamal1}
  J. Jalilian-Marian and A. H. Rezaeian, Phys. Rev. {\bf D85}, 014017 (2012) [arXiv:1110.2810]. 

\bibitem{jm}
J. L. Albacete and C. Marquet, Phys. Lett. {\bf B687}, 174 (2010).  

\bibitem{raj}
K. Dusling, F. Gelis, T. Lappi and R. Venugopalan, Nucl. Phys. {\bf A836}, 159 (2010) [arXiv:0911.2720]. 

\bibitem{bk-b}
K. Golec-Biernat and A. M. Stasto, Nucl. Phys. {\bf B668}, 345 (2003); J. Berger and  A. M. Stasto, Phys. Rev. {\bf D84}, 094022 (2011); arXiv:1205.2037.

\bibitem{urs}
N. Armesto, C. A. Salgado, and U. A. Wiedemann, Phys. Rev. Lett. {\bf 94},  022002 (2005). 

\bibitem{ncoll}
D. d'Enterria, nucl-ex/0302016. 

\bibitem{mstw}
A. D. Martin, W. J. Stirling, R. S. Thorne and G. Watt, Phys. Lett. {\bf B652}, 292 (2007);
A. D. Martin, W. J. Stirling, R. S. Thorne and G. Watt, Eur. Phys. J. {\bf C63}, 189 (2009). 

\bibitem{kkp}
B. A. Kniehl, G. Kramer and B. Potter, Nucl. Phys. {\bf B582}, 514 (2000).

\bibitem{ffp}
L. Bourhis, M. Fontannaz and J. P. Guillet, Eur. Phys. J. {\bf C2}, 529 (1998); M. Gluck, E. Reya and A. Vogt, Phys. Rev. {\bf D48}, 116 (1993), Erratum-ibid. {\bf D51}, 1427 (1995). 
\bibitem{dy}
A. Stasto, B-W Xiao and D. Zaslavsky, Phys. Rev. {\bf D86}, 014009 (2012). 

\bibitem{ridge-cms}
V. Khachatryan {\it et al.} [CMS Collaboration], JHEP {\bf 1009}, 091 (2010).  
\bibitem{alex}
A. Kovner and M. Lublinsky, Phys. Rev. {\bf D83}, 034017 (2011); {\bf D84}, 094011 (2011). 
 






\end{thebibliography}
\end{document}